\documentclass[usenatbib]{mn2e}
\usepackage{rotating}
\usepackage{epsfig}
\usepackage{times}
\def\gtsim{~\rlap{$>$}{\lower 1.0ex\hbox{$\sim$}}}
\def\ltsim{~\rlap{$<$}{\lower 1.0ex\hbox{$\sim$}}}

\title[Flux calibration of the Herschel-SPIRE photometer] 
    {Flux calibration of the {\it Herschel}\thanks{Herschel 
    is an ESA space observatory with science instruments provided by 
    European-led Principal Investigator consortia and with important
    participation from NASA.}-SPIRE photometer}

\author[G. J. Bendo et al.]
    {G. J. Bendo$^1$, M. J. Griffin$^2$, J. J. Bock$^3$, L. Conversi$^4$, 
          C. D. Dowell$^3$, T. Lim$^5$, N. Lu$^6$,\newauthor
     C. E. North$^2$, A. Papageorgiou$^2$, C. P. Pearson$^{5,7}$, M. Pohlen$^8$, 
          E. T. Polehampton$^5$,\newauthor
     B. Schulz$^6$, D. L. Shupe$^6$, B. Sibthorpe$^{9,10}$, L. D. Spencer$^2$, 
          B. M. Swinyard$^{5,11}$,\newauthor
     I. Valtchanov$^4$, C. K. Xu$^6$\\
    $^1$   UK ALMA Regional Centre Node, Jodrell Bank Centre for Astrophysics, 
           School of Physics and Astronomy, University of Manchester, 
           Oxford Road,\\ Manchester M13 9PL, United Kingdom\\
    $^2$   School of Physics and Astronomy, Cardiff University, 
           Queens Buildings, The Parade, Cardiff CF24 3AA, United Kingdom\\
    $^3$   NASA Jet Propulsion Laboratory, 4800 Oak Grove Drive, Pasadena,
           CA 91109, USA\\
    $^4$   Herschel Science Centre, ESAC, ESA, PO Box 78, Villanueva de la 
           Ca\~nada, 28691 Madrid, Spain\\
    $^5$   Space Science and Technology Department, Rutherford Appleton 
           Laboratory, Chilton, Didcot, Oxfordshire OX11 0QX, United Kingdom\\
    $^6$   NASA Herschel Science Center, IPAC, 770 South Wilson Avenue,
           Pasadena, CA 91125, USA\\
    $^7$   Department of Physical Sciences, The Open University, 
           Milton Keynes MK7 6AA, UK\\
    $^8$   Gemini Observatory, Northern Operations Center, 670 N. A’ohoku 
           Place, Hilo, HI 96720, USA\\
    $^9$   UK Astronomy Technology Centre, Royal Observatory Edinburgh, 
           Blackford Hill, Edinburgh EH9 3HJ, United Kingdom\\
    $^{10}$ SRON Netherlands Institute for Space Research, NL-9747 AD 
           Groningen, 
           The Netherlands\\
    $^{11}$ Department of Physics and Astronomy, University College London, 
           Gower Street, London WC1E 6BT, United Kingdom\\ 
    }
\date{}
\pagerange{\pageref{firstpage}--\pageref{lastpage}}
\pubyear{}

\begin{document}
\label{firstpage}
\maketitle

\begin{abstract}
We describe the procedure used to flux calibrate the three-band
submillimetre photometer in the Spectral and Photometric Imaging
REceiver (SPIRE) instrument on the {\it Herschel} Space Observatory.
This includes the equations describing the calibration scheme, a
justification for using Neptune as the primary calibration source, a
description of the observations and data processing procedures used to
derive flux calibration parameters (for converting from voltage to
flux density) for every bolometer in each array, an analysis of the
error budget in the flux calibration for the individual bolometers,
and tests of the flux calibration on observations of primary and
secondary calibrators.  The procedure for deriving the flux
calibration parameters is divided into two parts.  In the first part,
we use observations of astronomical sources in conjunction with the
operation of the photometer internal calibration source to derive the
unscaled derivatives of the flux calibration curves.  To scale the
calibration curves in Jy beam$^{-1}$ V$^{-1}$, we then use
observations of Neptune in which the beam of each bolometer is mapped
using Neptune observed in a very fine scan pattern.  The total
instrumental uncertainties in the flux calibration for most individual
bolometers is $\sim0.5$\%, although a few bolometers have
uncertainties of $\sim1-5$\% because of issues with the Neptune
observations.  Based on application of the flux calibration parameters
to Neptune observations performed using typical scan map observing
modes, we determined that measurements from each array as a whole have
instrumental uncertainties of 1.5\%.  This is considerably less than
the absolute calibration uncertainty associated with the model of
Neptune, which is estimated at 4\%.
\end{abstract}


\section{Introduction}

The Spectral and Photometric Imaging REceiver \citep[SPIRE;
][]{getal10}, is one of the three instruments on the {\it Herschel}
Space Observatory \citep{petal10}. It has separate photometric and
spectroscopic imaging sub-instruments that take advantage of {\it
Herschel's} capabilities for submillimetre observations.  The
photometer has three individual arrays of feedhorn-coupled bolometers
using Neutron Transmutation Doped (NTD) germanium thermistors
\citep{tetal01} cooled to approximately 0.3 K by an internal helium-3
refrigerator.  The arrays cover three broad passbands centered at
approximately 250, 350 and 500~$\mu$m and contain 139, 88, and 43
bolometers, respectively.  The photometer is primarily used in a scan
map mode that produces images in a single observation ranging from
$4\times4$ arcmin to several square degrees.

A flux calibration method appropriate for broadband submillimetre
photometric instruments is described in Griffin et al. (2013,
submitted).  This method involves the conversion of linearised
bolometer voltage signals to monochromatic flux densities or sky
surface brightness values using measurements of a calibration standard
and knowledge of the relevant instrument properties.  The purpose of
this paper is to describe the detailed implementation of this flux
calibration scheme in the case of the SPIRE photometer.  The resulting
parameters describing the calibration curves are used by the flux
conversion module within the SPIRE photometer data processing pipeline
\citet{getal08, detal10}.

Section~\ref{s_equations} gives an overview of the equations
(including non-linearity corrections) describing the conversion of the
measured bolometer output voltage to flux density.
Section~\ref{s_calstandard} provides information on the use of Neptune
as the primary flux standard, including the calculation of the Neptune
flux density.  Descriptions of the observations that were performed to
derive the terms in the flux calibration equations are given in
Section~\ref{s_obsoverview}, and Sections~\ref{s_curve} and
\ref{s_scale} provide details on the analysis used to derive the
calibration parameters for every bolometer in the SPIRE photometer
arrays.  Separate calibration terms are derived for the two standard
bias voltage settings used for photometer observations: the nominal
settings (used for most observations) which are optimised for sources
fainter than 200~Jy beam$^{-1}$, and the bright source settings, which
are intended to be used for sources brighter than 200~Jy beam$^{-1}$.
A discussion of problematic bolometers and how they were handled is
given in Section~\ref{s_probbolom}.  Sources of uncertainty in the flux
calibration curves are discussed in Section~\ref{s_unc}, while
Section~\ref{s_test} describes tests of the derived flux calibration
curves using observations of primary and secondary calibration
sources.  Section~\ref{s_summary} provides a detailed summary of the
results from the assessment of the flux calibration.  Throughout this
paper, individual bolometers are referred to by their SPIRE array
designations PSW, PMW, and PLW for Photometer (Short, Medium, and Long)
Wavelength (with the wavelength corresponding to 250, 350, and
500~$\mu$m, respectively) followed by row letter and column number (e.g.,
PSWE2).

\section{Conversion of bolometer voltage to flux density}
\label{s_equations}

The SPIRE detectors, being bolometers, respond not to the absorbed
photon rate but to the amount of power that they absorb.  As noted by
Griffin et al. (2013, submitted), the absorbed power is proportional
to the spectral response function (SRF) weighted flux density, which
is the flux density weighted by the overall SRF and integrated across
the passband.  This is given by
\begin{equation}
\bar{S}_{Meas} = \frac{\int_{\nu}S(\nu)F(\nu)\eta(\nu)d\nu}
  {\int_{\nu}F(\nu)\eta(\nu)d\nu},
\label{e_srffluxdensity}
\end{equation}
where $F(\nu)$ is the SRF and $\eta(\nu)$ is the aperture efficiency.
The SPIRE photometer SRFs, which were measured by Fourier transform
spectroscopy, are shown in Figure 7 of Griffin et al. (2013,
submitted) and in the SPIRE Observers’ Manual
\citep{spire11}\footnote[12]{The SPIRE Observers' Manual is available
  at http://herschel.esac.esa.int/Docs/SPIRE/pdf/spire\_om.pdf .}, and
they are also available in tabular form within the Herschel
Interactive Processing Environment \citep[HIPE; ][]{o10}.

For NTD bolometers, the small-signal responsivity (variation of output
voltage with absorbed radiant power) depends on the total voltage
across the bolometer in a manner which is approximately linear over a
wide range of background loading and bath temperature conditions
\citep{g07}\footnote[13]{This document is available at
  http://herschel.esac.esa.int/twiki/pub /Public/SpireCalibrationWeb/SPIRE\_Detector\_Parameter\_Sensitivity\_Issue \_1\_Nov\_14\_2007.pdf .}
The relation between a small change in the in-beam SRF-weighted flux
density, $d\bar{S}_{Meas}$, and the corresponding change in the
voltage across the bolometer, $dV$, can be expressed as
\begin{equation}
\frac{d\bar{S}_{Meas}}{dV} = f(V)
\end{equation}
where $f(V)$ is allowed to depart from linearity.  We find that the
differential responsivity for the SPIRE bolometers can be well
represented by
\begin{equation}
f(V)=K_1+\frac{K_2}{V-K_3},
\label{e_fdefinition}
\end{equation}
as is demonstrated in Section \ref{s_curve_derivation}.  In this
equation, $K_1$, $K_2$, and $K_3$ are constants specific to each
bolometer in each bias voltage setting.  $K_1$ has units of Jy
V$^{-1}$, $K_2$ has units of Jy, and $K_3$ has units of V. The
conversion between a measured voltage $V_m$ and the corresponding
SRF-weighted flux density is obtained from the integral of $f(V)$
given by
\begin{equation}
\bar{S}_{Meas} = \int^{V_m}_{V_0}f(V)dV,
\end{equation}
where $V_0$ is the operating point voltage (the signal that would be
measured when viewing dark sky) and $V_m$ is the measured voltage.
The result of this integral is
\begin{equation}
\bar{S}_{Meas} = K_1 (V_m-V_0) + K_2 \ln \left( \frac{V_m-K_3}{V_0-K_3} 
  \right).
\label{e_cal}
\end{equation}

The SPIRE calibration scheme is thus based on deriving the
$K$-parameters that describe $f(V)$.  For calibration of $f(V)$, the
photometer internal calibration source \citep[PCal; ][]{petal05} is
used to provide a repeatable small change in power illuminating the
detector.  The inverse of the corresponding detector response, $\Delta V_P$,
for a given bolometer voltage, $V$, is directly proportional to $f(V)$.
The relation can be written as
\begin{equation}
\frac{1}{\Delta V_P(V)} = Af(V) =  AK_1+\frac{AK_2}{V-K_3},
\end{equation}
where $A$ is a constant.  The shape of the function $f(V)$ can thus be
determined by observing changes in the signal from uniform PCal
flashes while varying the effective bolometer operating point voltage
$V$.  This could be done by changing the temperature of the helium-3
bath, but for SPIRE, it is more straightforward to vary $V$ by viewing
regions with different in-beam flux densities.  It is important to
note that in this exercise that the flux density does not actually
need to be known; the purpose of exposing the detector to a range of
source brightnesses is merely to vary the operating point voltage of
the detectors over the range of interest so as to characterise $f(V)$.

Having determined the shape of $f(V)$ from these PCal measurements,
the absolute value of the constant $A$ can be found from observations
of a source with a known SRF-weighted flux density, which then allows
for determining the numerical values of the $K$-parameters.  Following
Equation~\ref{e_srffluxdensity}, the SRF-weighted flux density for a
calibration source with a spectrum $S_C(\nu)$ is given by
\begin{equation}
\bar{S}_C = K_{Beam}(\nu)\left[\frac{\int_{\nu}S_C(\nu)F(\nu)\eta(\nu)d\nu)}
  {\int_{\nu}F(\nu)\eta(\nu)d\nu}\right],
\end{equation}
where $K_{Beam}(\nu)$ is a correction factor for possible partial
resolution of the calibrator by the telescope beam.

The equation relating the SRF-weighted flux density of the calibration
source, $\bar{S}_{C}$, to the scaling term $A$ can be written as
\begin{equation}
A=\frac{1}{\bar{S}_{C}} \int^{V_{ON}}_{V_{OFF}} \frac{1}{\Delta V_P(V)}dV
\end{equation}
where $V_{ON}$ is the voltage measured when the bolometer is pointed
at the calibrator and $V_{OFF}$ is the voltage measured off-source.

After determining the $K$-parameters in this way, the measured
SRF-weighted flux density for an unknown source can be found from the
corresponding measured bolometer voltage.  Conversion of the
SRF-weighted flux density to a monochromatic flux density requires a
choice of the frequency $\nu_{0}$ at which the flux density is to be
calculated, knowledge of the instrument SRF, an assumption concerning
the source spectral shape, and the stipulation that the source and
calibrator have the same spatial characteristics to ensure that the
conversion from in-beam flux density to absorbed detector power is the
same for the source and calibration observations.  When adopting a
point-like source as the calibrator, as is the case for SPIRE, it is
also appropriate to quote source flux densities based on the
assumption that the source is also point-like.

The convention adopted for {\it Herschel} is to quote monochromatic
flux density values under the assumption that $\nu S(\nu$) is constant
(i.e. the spectral index $\alpha$ is -1).  The conversion between the
SRF-weighted flux density and the monochromatic flux density
$S(\nu_0)$ is given by Griffin et al. (2013, submitted) as
\begin{equation}
S(\nu_0)=K_{MonP}(\alpha,\nu_0)\bar{S}_{Meas}, 
\end{equation}
where $K_{MonP}$ is computed for $\alpha=-1$ and the chosen value of
$\nu_{0}$ using
\begin{equation}
K_{MonP}(\alpha,\nu_0) = \frac{\int_{\nu} F(\nu) \eta(\nu)
    d\nu}{\int_{\nu} \left( \frac{\nu}{\nu_0} \right) ^\alpha F(\nu)
    \eta(\nu) d\nu}.
\end{equation}
Values for $K_{MonP}$ and the wavelengths $\lambda_o$ corresponding to
$\nu_o$ are given in Table~\ref{t_kmonp}.

\begin{table}
\caption{Values of $K_{MonP}$}
\label{t_kmonp}
\begin{center}
\begin{tabular}{@{}lcc@{}}
\hline
Array &    $\lambda_0$ &    $K_{MonP}$\\
&          ($\mu$m) &       \\
\hline
PSW &      250 &            1.0102 \\
PMW &      350 &            1.0095 \\
PLW &      500 &            1.0056 \\
\hline
\end{tabular}
\end{center}
\end{table}

For SPIRE, the flux densities produced by the data pipeline are quoted
at frequencies corresponding to standard wavelengths of 250, 350, and
500~$\mu$m.  The emission from most sources observed by SPIRE is from
dust, and the continuum spectrum is not well described by a spectral
index of -1.  The derivation of the necessary colour correction
functions to account for the different spectral shapes is described by
Griffin et al. (2013, submitted), and the functions for SPIRE are
given in the SPIRE Observers' Manual \citep{spire11}.  The
  accurate calibration of extended and semi-extended emission also
  requires a detailed knowledge of the beam profile and aperture
  efficiency, and this is also discussed by Griffin et al. (2013,
  submitted) and in the SPIRE Observer's Manual.

\section{Primary calibration standard (Neptune)}
\label{s_calstandard}

\begin{figure*}
\begin{center}
\epsfig{file=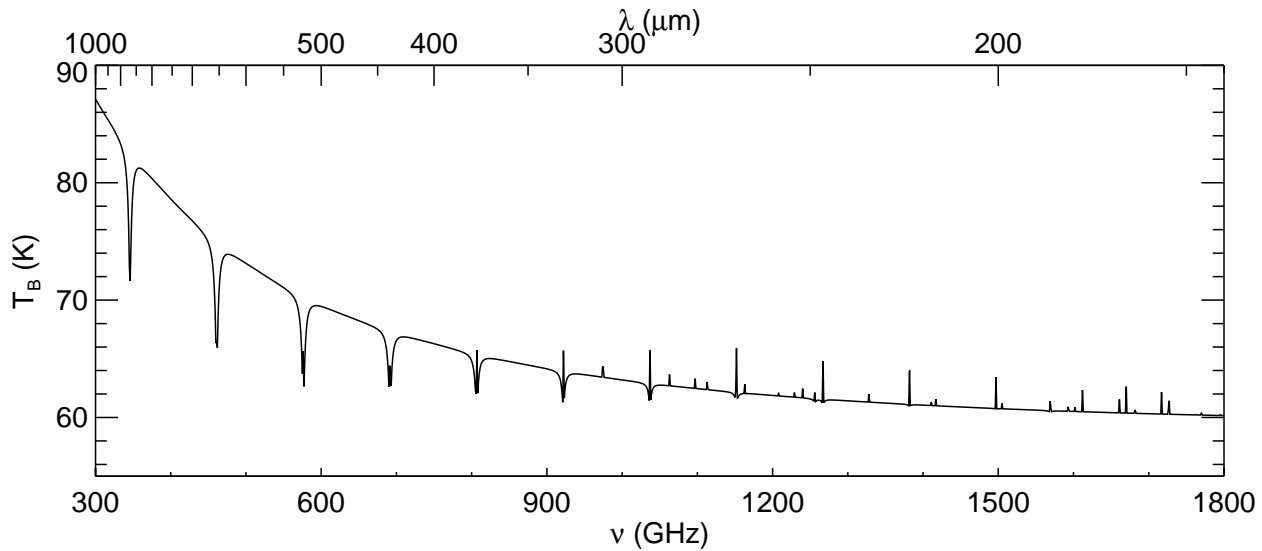}
\caption{The disk-averaged model brightness temperature of Neptune as a 
function of frequency and wavelength.}
\label{f_tbnep}
\end{center}
\end{figure*}

The planet Neptune is used as the primary calibration source for the
SPIRE photometer.  It has a well-understood submillimetre spectrum, it
is almost completely point-like in the SPIRE beams (the angular
diameter of Neptune is $\sim2$~arcsec, while the 250, 350, and
500~$\mu$m arrays have full-width at half-maxima (FWHM) of 18, 25, and
36~arcsec, respectively), and it is sufficiently bright to provide
high S/N when observed but not so bright as to introduce any
significant non-linear response from the detectors. We adopt a model
of the disk-averaged brightness temperature spectrum of Neptune based
on the ESA-4 version of the planetary atmosphere model first published
by \citet{m98}\footnote[14]{The ESA-4 models for Uranus and Neptune
  are available at
  ftp://ftp.sciops.esa.int/pub/hsc-calibration/PlanetaryModels/ESA4/
  .}.  The absolute photometric uncertainty of this model is $\pm4$\%
(R. Moreno, private communication), with the true brightness
temperature being enclosed within those bounds.  The absolute
uncertainty is mainly attributed to the uncertainty in molecular
absorption coefficients ($\sim1$\%) and the adopted temperature
structure ($\sim3$\%).  The Neptune brightness temperature spectrum in
the SPIRE range is shown in Figure~\ref{f_tbnep}.  The absorption
features are caused by CO, and the emission lines are caused by HCN.

To calculate the solid angle of Neptune for a given observation we use
an equatorial radius, $r_{eq}$ of 24,766 km and polar radius $r_{p}$
of 24,342 km, both of which are based on the analysis of Voyager data
by \citet{l92}.  These are similar to the values used in the
ground-based observations by \citet{hetal85}, \citet{oetal86}, and
\citet{go93}.

In calculating the planetary angular sizes and solid angles, a
correction is applied for the inclination of the planet's axis at the
time of observation.  The apparent polar radius $r_{p-a}$ is given by
\citet{m1897} as
\begin{equation}
r_{p-a}=r_{eq}\left[ 1-e^2 \cos ^2 (\phi) \right]^\frac{1}{2}.
\end{equation}
In this equation, $\phi$ is the latitude of the sub-{\it Herschel} point, and
$e$ is the planet's eccentricity, which can be calculated using
\begin{equation}
e=\left[ \frac{r_{eq}^2-r_p^2}{r_p^2}\right]^\frac{1}{2}.
\end{equation}
The observed planetary disc is taken to have a geometric mean radius,
$r_{gm}$, given by
\begin{equation}
r_{gm}=\left[r_{eq}r_{p-a}\right]^\frac{1}{2}.
\end{equation}

\begin{figure*}
\begin{center}
\epsfig{file=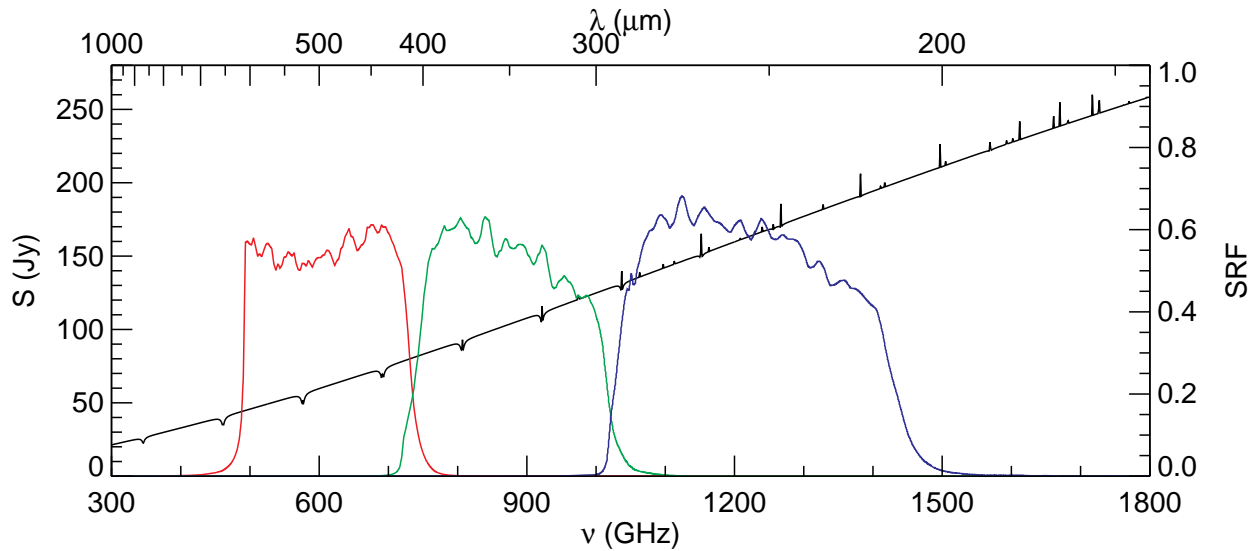}
\caption{The model spectrum of Neptune on OD 168 (29 Oct 2009) in
  black along with the SRFs of the three SPIRE photometer bands (from 
  ftp://ftp.sciops.esa.int/pub/hsc-calibration/SPIRE/PHOT/Filters/), with
  the blue, green, and red curves corresponding to the 250, 350, and
  500~$\mu$m bands, respectively.}
\label{f_trans}
\end{center}
\end{figure*}

The {\it Herschel}-planet distance, is obtained from the NASA JPL
Horizons ephemeris system \citep{getal96}\footnote[15]{The ephemeris
  can be accessed at http://ssd.jpl.nasa.gov/?horizons .}, and the
observed angular radius and solid angle are calculated
accordingly. The Neptune flux density spectrum, $S_C(\nu)$ at the {\it
  Herschel} telescope aperture is computed from the solid angle and
the disk-averaged brightness temperature spectrum.  Over a period of
one year, the brightness of Neptune exhibits $\pm7$\% variations due
to the seasonally varying {\it Herschel}-Neptune distance.  However,
because Neptune is only observable when it crosses through one of two
visibility windows, Neptune varied by only $\pm2$\% in brightness when
it was observable by {\it Herschel}.  Typical flux densities at the
nominal SPIRE wavelengths are 160, 100, and 60 Jy at 250, 350, and 500
$\mu$m, respectively. Figure~\ref{f_trans} shows the model spectrum of
Neptune on {\it Herschel} operational day (OD) 168 along with the SRFs
of the three SPIRE photometer bands. The slope of the Neptunian
submillimetre continuum is somewhat less steep than that of a black
body because of the increase in brightness temperature with decreasing
frequency, with lower frequencies probing deeper and warmer parts of
the troposphere.  The continuum spectral indices ($\alpha$ as given by
$S(\nu) \propto \nu^\alpha$, averaged across the bands) are 1.26,
1.39, and 1.45 for the 250, 350, and 500~$\mu$m bands, respectively.

\section{Overview of derivation of calibration terms}
\label{s_obsoverview}

Following the procedure outlined in Section~\ref{s_equations}, the
derivation of the flux calibration was broken into two steps. The
first step was to identify the shape of $f(V)$ as given by
Equation~\ref{e_fdefinition}.  In these observations, the telescope
stared at a series of regions with different surface brightness values
(thus providing a range of in-beam flux densities).  With the
stationary telescope position providing a fixed photometric
background, and therefore a fixed operating point voltage, PCal
flashes were applied to generate a small additional modulation of the
detector signal.  This gave both a measurement of the voltage ($V$)
and of the change in voltage relative to a small change in signal
($d\bar{S}_{Meas}/dV$).  For the reason explained in
Section~\ref{s_equations}, it was important that the observations
sample a series of regions with widely varying surface brightnesses to
provide a sufficiently wide range of operating voltages.  We used
observations of bright extended emission in Sgr A as well as several
nearby background regions to derive the unscaled calibration curves
for the nominal bias detector settings (which are used for almost all
observations).  For the bright source mode, which is used for only a
few sources brighter than $\sim200$ Jy beam$^{-1}$, we used pointings
near Sgr B2.  Further details on the observations and data analysis
are given in Section~\ref{s_curve}.

Observations of the primary calibration source Neptune by each and
every bolometer were needed to set the absolute scale for the
calibration parameters.  We observed the planet using a special ``fine
scan'' map mode in which every bolometer was scanned over the target
with a finely-spaced grid.  We then fitted two-dimensional Gaussian
functions to the timeline data for each bolometer to determine the
peak voltage (the difference between the on-target voltage
  $V_{ON}$ and the off-target voltage $V_{OFF}$) measured by each
bolometer when centered on Neptune.  These peak voltages were then
used along with the Neptune model flux densities to scale the flux
calibration curves.  Details are given in Section~\ref{s_scale}.

The measured flux density depends on the value of $V_0$ that is
used.  Ideally $V_0$, which corresponds to the bolometer voltage when
viewing blank sky, would be a constant over the life of the mission.
However most of the radiant power incident on the SPIRE bolometers is
from the warm telescope, and because the telescope temperature varies
by several K depending on the season and the solar aspect angle, unique
values of $V_0$ cannot be defined.  For the SPIRE pipeline we have
adopted a nominal set of values based on dark sky observations early
in the mission and with a representative telescope temperature.  The
$V_0$ values for the nominal mode were derived for each bolometer from
the median voltages measured in dark sky observation 1342182454, which were
performed during Operational Day (OD) 98 (19 Aug 2009), and the $V_0$
values for the bright source mode were derived from data taken during
observation 1342185829 on OD 153 (14 Oct 2009).  SPIRE data taken in
other observations normally have slightly different temperatures and
therefore may have photometric offsets, so SPIRE data alone cannot be used
to make absolute estimates of the sky intensity.  These offsets are
ultimately subtracted off in the mapmaking procedure.

\section{Defining the shape of unscaled flux calibration curve}
\label{s_curve}

\subsection{Observing procedure}

\begin{figure*}
\begin{center}
\epsfig{file=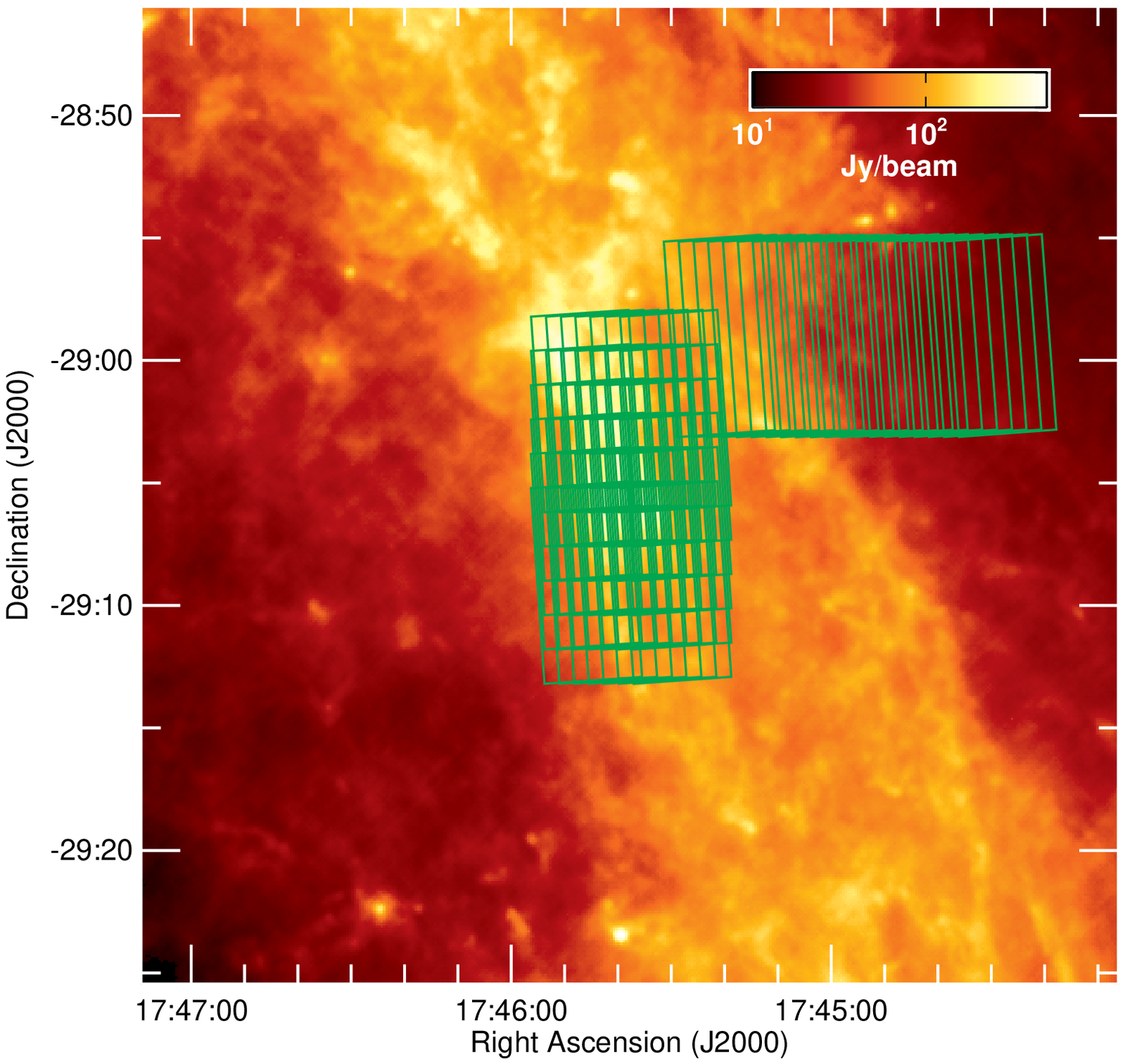}
\caption{The SPIRE 250~$\mu$m image of Sgr A$^*$ from {\it Herschel}
  with green $8\times4$~arcmin regions overlaid to show the locations
  where the SPIRE arrays pointed during the nominal voltage mode
  observations with PCal flashes.  The image was created with data
  from parallel-mode scan map observations 1342204102 and 1342204103
  that were processed using a modified version of the standard
  parallel data processing pipeline and the flux calibration
  parameters that were derived in this paper.  The displayed area is
  $40\times40$~arcmin with north up and east to the left.}
\label{f_sgra}
\end{center}
\end{figure*}

To determine the unscaled versions of the $K$ parameters, we performed
a series of staring observations on fields with varying surface
brightness levels as PCal flashes are applied.  During each 1 min
observation, PCal was alternately turned on for 1.5 sec and off for
1.5 sec for a total of 20 cycles.  For the nominal detector settings,
the targets for these pointings consisted of a grid of locations
centered on Sgr A and an additional strip of pointings running roughly
perpendicular to the plane of the Galaxy as shown in
Figure~\ref{f_sgra}.  The locations were selected using estimates of
the expected surface brightness in the SPIRE bands based on the
Caltech Submillimeter Observatory 350~$\mu$m data from
\citet{babetal10}.  The pointings around Sgr A provided data for the
high signal end of the calibration curve, while the strip to the
northwest provided data for lower signals.  The observations were
planned taking into account that, when the target was visible, the
$4\times8$~arcmin SPIRE arrays were oriented so that the long (y) axis
was approximately aligned with right ascension and the short (z) axis
aligned with declination.  A key requirement of these observations was
to point every detector at least once at the region where the signal
is $\gtsim75$\% of the peak surface brightness of Sgr A.  This region
is roughly $0.75 \times 2$~arcmin in size with its major axis aligned
slightly counter-clockwise of north-south.  Based on simulating
observations of this region, we adopted a grid of observations
consisting of 7 columns spaced by 36~arcsec(the distance in the z
direction between two rows in the PMW array) and 6 rows spaced by
42~arcsec.  In the strip of pointings to the north of Sgr A, we used
20 pointings spaced by 36~arcsec.  In the total set of observations,
the surface brightness from the background measured by each bolometer
was predicted to vary by a factor of $\sim10$ and to approximately
equal or exceed the surface brightness of Neptune at the high end of
the range.

\begin{figure*}
\begin{center}
\epsfig{file=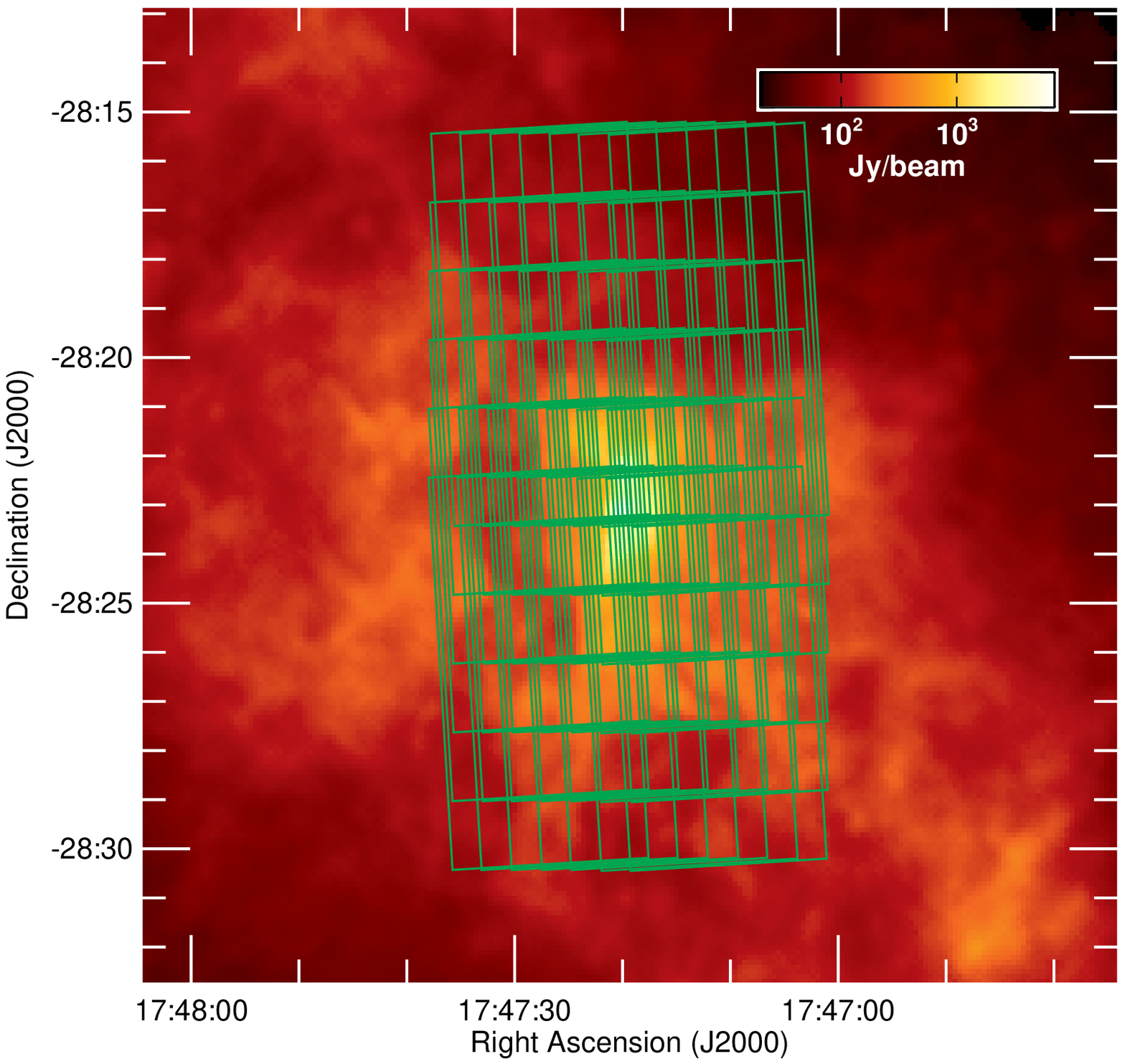}
\caption{The SPIRE 250~$\mu$m image of Sgr B2 from {\it Herschel} with
  green $8\times4$~arcmin regions overlaid to show the locations where
  the SPIRE arrays pointed during the bright source mode observations
  with PCal flashes.  The image was created with data from
  parallel-mode scan map observations 1342184474 and 1342184475 that
  were processed using a modified version of the standard parallel
  data processing pipeline and the flux calibration parameters that
  were derived in this paper.  The displayed area is
  $20\times20$~arcmin with north up and east to the left.}
\label{f_sgrb2}
\end{center}
\end{figure*}

For the bright detector settings, the targets consisted of a grid of
locations centered on Sgr B2 as shown in Figure~\ref{f_sgrb2}.  These
locations were also selected using estimates of the expected surface
brightness in the SPIRE wavebands based on CSO 350~$\mu$m data from
\citet{babetal10}.  We took into account that, when Sgr B2 was
visible, the array would be oriented so that the long axis would be
aligned with right ascension and the short axis would be aligned with
declination, just as was the case with Sgr A.  As with the Sgr A
observations, it was necessary to point every detector at least once
at the $0.6 \times 1.4$~arcmin region with a surface brightness that
is at least 75\% the peak surface brightness of Sgr B2.  We used a
grid of pointings identical to that used for Sgr A.  During these
observations, the in-beam flux density was expected to vary by a
factor of 35-40 for the typical bolometer in each array.

These observations of Sgr A and Sgr B2 were performed on OD 153 (14
Oct 2009).  In addition, we also used all other usable calibration
data taken between OD 116 (07 Sep 2009; the first date when usable
observations with PCal flashes had been performed after the current
bias voltage levels were set for the nominal and bright source modes)
and OD 424 (12 Jul 2010; the last date when observations with PCal
flashes were performed).  This included observations of dark sky and
observations in which point-like calibration sources (Mars, Uranus,
Neptune, 3 Juno, 4 Vesta, Alpha Boo, Gamma Dra, and VY CMa) were
observed for various calibration tests.  These data were used to
constrain the low signal end of the calibration curves.

\subsection{Derivation of the unscaled flux calibration curve}
\label{s_curve_derivation}

\begin{figure*}
\begin{center}
\epsfig{file=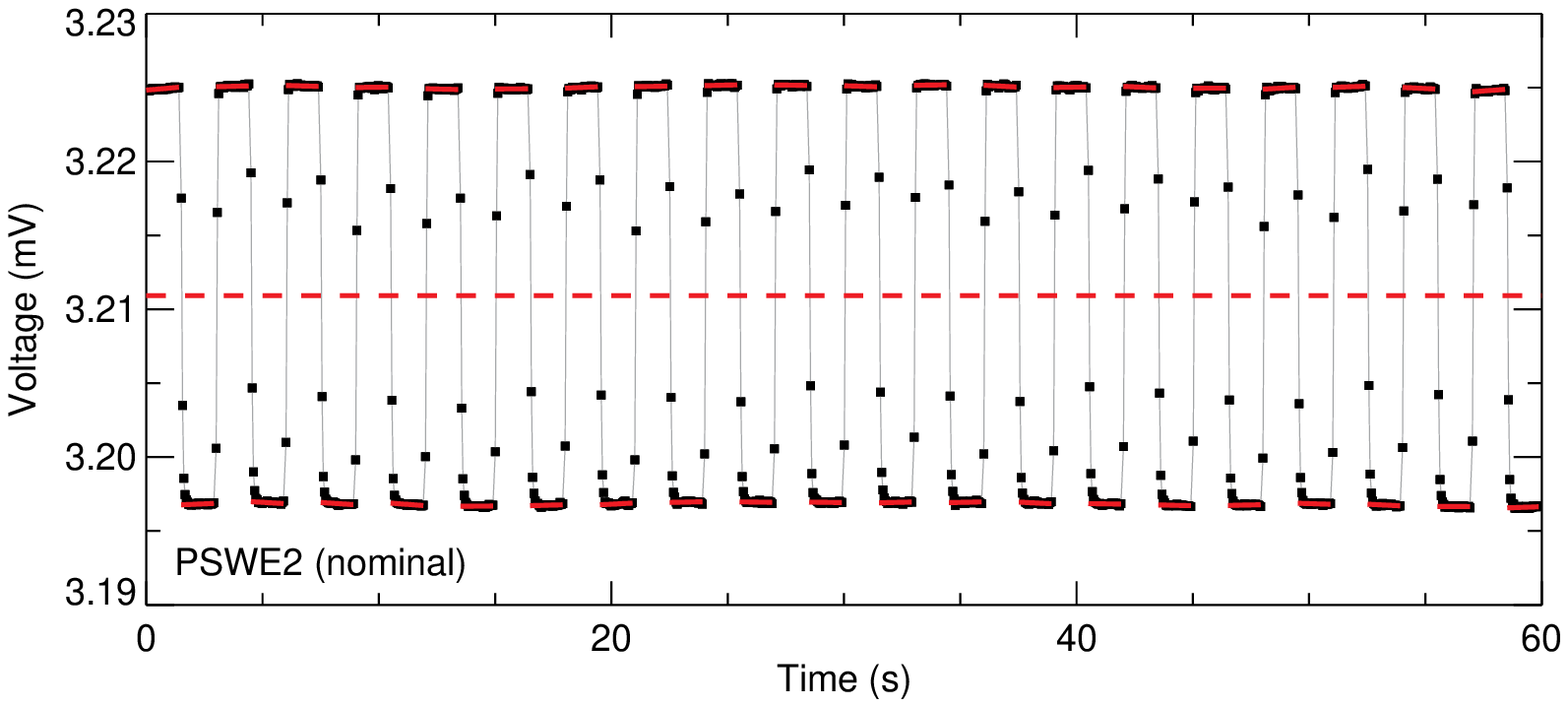}
\caption{An example of an observation with PCal flashes in the nominal
  voltage mode.  These are data taken by bolometer PSWE2 in a staring
  observation of Sgr A* with ID 1342185872.  The black squares show
  the individual measurements; the grey line joining the squares is
  used to aid in the visualisation of how the signal varied over time.
  Higher voltage values correspond to the signal measured when PCal was
  off, while lower voltage values correspond to the signal observed
  when PCal was on.  The dashed red line shows the mean voltage
  measured during the observations ($(3.2109 \pm
  0.0004)\times10^{-3}$~V), and the solid red lines show the functions
  fitted to the timeline segments when PCal was on or off.  The
  $\Delta V$ for these data was measured as $-(2.821 \pm
  0.003)\times10^{-5}$ V.}
\label{f_examplepcal_nominal}
\end{center}
\end{figure*}

\begin{figure*}
\begin{center}
\epsfig{file=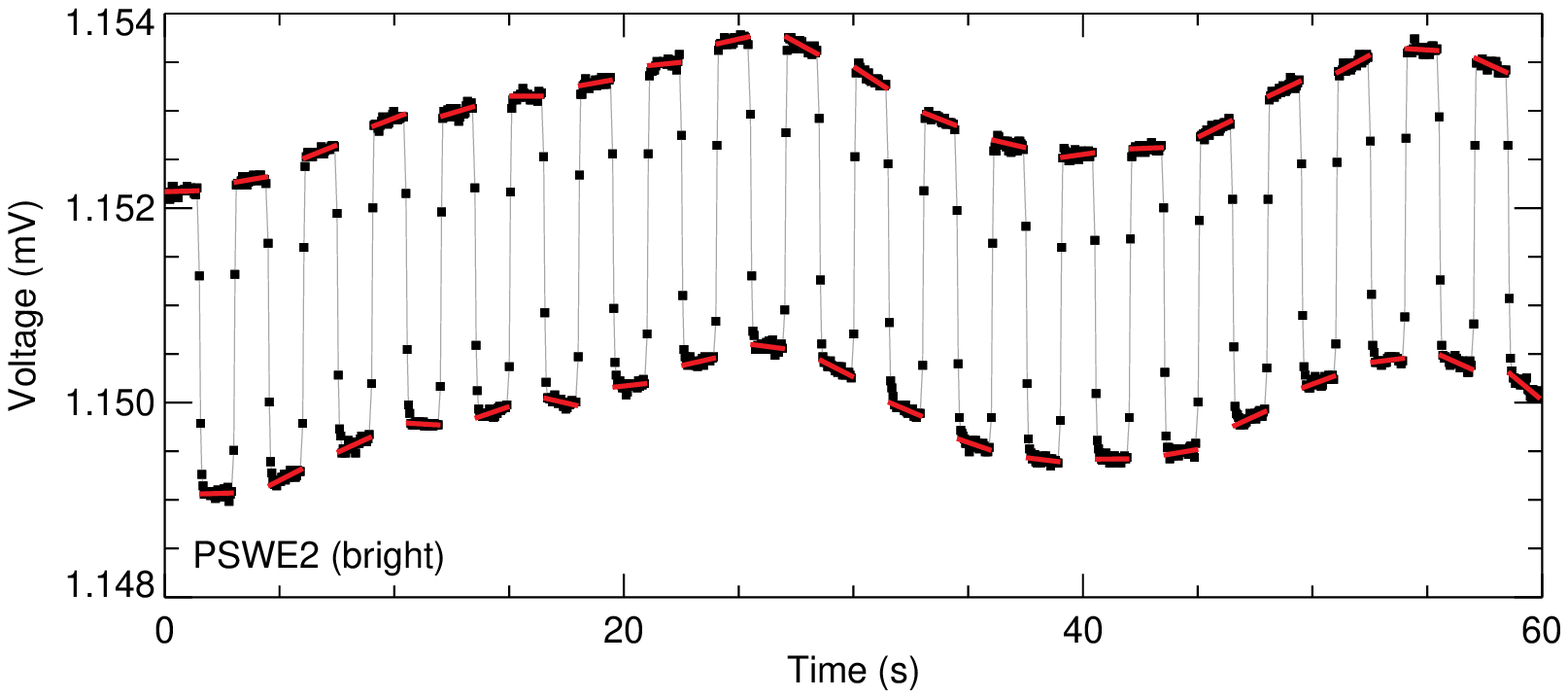}
\caption{An example of an observation with PCal flashes in the bright
  source mode.  These are data taken by bolometer PSWE2 in a staring
  observation of Sgr B2 with ID 1342185949.  The black squares show
  the individual measurements; the grey line joining the squares is
  used to aid in the visualisation of how the signal varied over time.
  Higher voltage values correspond to the signal measured when PCal was
  off, while lower voltage values correspond to the signal observed
  when PCal was on.  The solid red lines show the functions fitted to
  each segment of the square wave during the observations.  The mean
  voltage and uncertainty measured during the observations, based on
  the measured voltages for each step in the signal, is $(1.1515 \pm
  0.0004)\times10^{-3}$~V).  The mean and uncertainty in the $\Delta
  V$ for these data were measured as $-(3.18 \pm 0.04)\times10^{-6}$
  V.}
\label{f_examplepcal_bright}
\end{center}
\end{figure*}

Examples of the PCal flash data obtained in the nominal and bright
source voltage modes are shown in
Figures~\ref{f_examplepcal_nominal} and
\ref{f_examplepcal_bright}.  In the nominal mode data, the PCal flash
provides a change in voltage which is large compared to the noise
level but represents a small change in $V$ (approximately a 0.03 mV
change to a 3.2 mV measurement).  In the bright source mode data, the
magnitude of the PCal flash may be equivalent to or smaller than the
variations in the background signal, as shown in the example.  This
occurred mainly in the centre of Sgr B2, which is very bright compared
to PCal but relatively compact.  Pointing jitter, even though at a low
level, resulted in variations in the background signal that were large
relative to PCal.

The techniques used to measure the change in signal $\Delta V$ at a
background signal $V$ for both the nominal and bright source voltage
modes were similar.  The steps described here were applied to the
signal from each bolometer in each observation with PCal flashes.  We
first fitted lines to each segment of the square wave.  The gaps
between the lines fitted to sequential steps in the square wave gives
us individual $\Delta V$ measurements.  We then calculated the mean
and standard deviation of $\Delta V$ from each observation for each
bolometer after removing statistical outliers (data more than
$5\sigma$ from the mean).  The corresponding $V$ values were obtained
in different ways for each voltage bias mode.  For the nominal mode
data, we measured the mean and standard deviation in $V$ using all of
the data for each bolometer in each observation, as the relatively
small variation in the signal induced by the PCal flash ultimately
resulted in a relatively low standard deviation in $V$.  In the case
of the bright source mode data, we used the midpoint between the steps
in the square wave as individual $V$ measurements and then calculated
the mean and standard deviation of $V$ for each bolometer in each
observation.

\begin{figure}
\epsfig{file=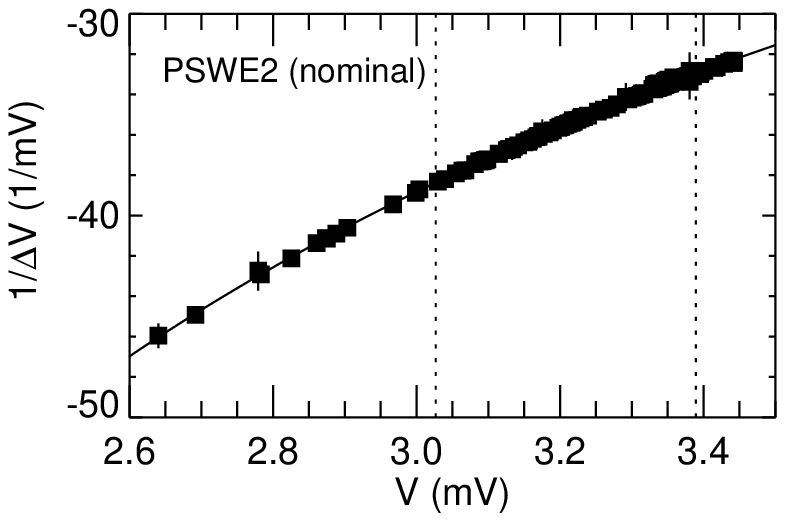}
\epsfig{file=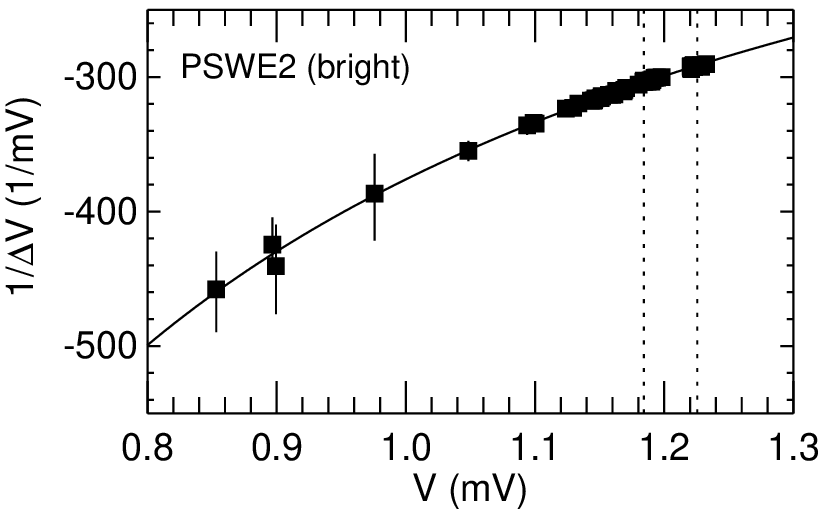}
\caption{PCal flash data for PSWE2.  Data are shown for both bias
  modes.  Lower voltages correspond to viewing brighter regions of
  sky.  The solid line shows the function described by
  Equation~\ref{e_fdefinition} that was fitted to the data.  The
  dotted lines show the mean peak and background signal measured for
  Neptune in the fine scan observations described in
  Section~\ref{s_scale}.}
\label{f_dcalcurve}
\end{figure}

We then fitted Equation~\ref{e_fdefinition} to the $V$ and $\Delta V$
data.  In the nominal voltage data, we excluded data with unusually
high uncertainties in $\Delta V$ ($>10^{-3}$~mV), which may originate
from instances where the prior statistical filtering steps have not
succeeded in removing data containing glitches.  Example plots,
including best fitting functions, are shown in
Figure~\ref{f_dcalcurve}.  The dotted lines show the average on-source
and off-source voltage levels measured for Neptune in the fine scan
observations described in Section~\ref{s_scale}.

Table ~\ref{t_pcalrange} gives the typical signal range in Jy
beam$^{-1}$ over which the calibration applies (calculated after
performing the scaling steps in Section~\ref{s_scale} first).  The
ranges were calculated in two steps.  First, we determined for each
bolometer the minimum and maximum signals measured during the
observations with PCal flashes.  Then for each array, we
determined the median of the maxima and the median of the minima.  The
flux calibration should be well-constrained within these signal
ranges, but measurements outside of these ranges should be treated
with caution.  Note that the range extends below 0 Jy beam$^{-1}$.
This is because the background signal in dark sky may sometimes drop
below below 0 as a result of thermal drift in the telescope and
instrument; minimum values below 0 arise from these variations in the
telescope background.  Users should attach no physical interpretation
to background signals below 0 measured in SPIRE data, as SPIRE is only
designed to measure signals relative to the background within fields
and not the absolute sky brightness.

\begin{table}
\caption{Representative signal ranges from zero point covered by
  PCal flash data (in Jy beam$^{-1}$)}
\label{t_pcalrange}
\begin{center}
\begin{tabular}{@{}lcccc@{}}
\hline
Array &      \multicolumn{2}{c}{Nominal Mode} &
             \multicolumn{2}{c}{Bright Source Mode}\\
&            Minimum &      Maximum &       Minimum &      Maximum\\
&            Values &       Values &        Values &       Values\\
\hline 
PSW &        -20 &          400 &           -30 &          2710 \\
PMW &        -20 &          360 &           -30 &          2200 \\
PLW &        -40 &          270 &           -50 &          1650 \\
\hline
\end{tabular}
\end{center}
\end{table}

\section{Scaling the flux calibration curve}
\label{s_scale}

\subsection{Observing procedure}

In the Neptune fine scan observations, each bolometer was scanned over
the planet in a series of parallel scan legs with $\sim2-3$~arcsec
offsets from each other.  Scans were performed in four directions: two
scans were aligned with the y-axis in the instrument plane, and two
scans aligned with the z-axis in the instrument plane.  Altogether,
the centre of Neptune passed within $\sim 1$~arcsec of each detector
four times.  The nominal voltage mode observations (observation
numbers 1342186522-1342186525) were performed on OD 168 (29 Oct 2009),
while the bright source mode observations (observation numbers
1342187438, 1342187439, 1342187507, and 1342187508) were performed on
OD 201-202 (01-02 Dec 2009).

\subsection{Measuring the peak signal in the timeline data}

To measure the peak signal from Neptune and the background signal, we
used a Levenberg-Marquardt algorithm to fit two-dimensional elliptical
Gaussian functions to the timeline data taken during each fine scan
observation by each bolometer.  The advantage that timeline-based beam
fitting has over map-based beam fitting is that the timeline data
contain the signal measurements as sampled at their precise pointed
positions whereas the map data are affected by smearing effects
related to pixelisation.  The limiting factor in the positional
accuracy of the timeline measurements is the relative pointing
uncertainty of the telescope, which is $\ltsim2$~arcsec in the data
used for this analysis \citep{petal10}.  In contrast, when the
timeline data are converted into maps, the signal from an individual
sample is assigned to a square pixel area (or, as is the case with
some mapmakers using drizzle methods, the signal is divided over
multiple pixels), so some spatial information is lost.  Moreover,
measurements passing across the edges of map pixels are effectively
shifted in position to the centres of the map pixels.  Hence,
mapmaking has the effect of smoothing the data, which suppresses the
peak signal from unresolved sources.  In using the timeline data, we
avoid these smoothing effects, thus allowing for more precise
measurements of the peak signals.

We performed beam fitting on the data from each fine scan observation
for each individual bolometer.  We first selected all data from all
bolometer samples that fell within a target radius and background
annular area based on central locations selected from map data.  The
position of the target aperture did not need to be accurately defined,
as tests with this method have demonstrated that the central position
can be offset from the target position by half of the beam FWHM and
still produce statistically similar results.  After the data for the
fit were selected, the central position of the source was treated as a
free parameter in the fit.  The target radius was set to extend to the
location of the minima between the peak and the first diffraction ring
of the beam profile, thus encompassing the central part of the
  beam that most closely resembles a Gaussian function.  The radii
were 22, 30, and 42~arcsec for the 250, 350, and 500~$\mu$m arrays,
respectively.  The background annulus was set to be 350 to 400~arcsec
in radius, which is large enough that signal from the outer
diffraction rings is not detectable within the area.  In the nominal
bias setting analysis, we used data from all scan legs that fell
within the target and background apertures.  In the bright source
mode, however, we found significant drift in the background signal,
which caused problems when attempting to fit the background.  In that
case, we used all data points that fell within the target aperture,
but we only selected data falling in the background aperture from scan
legs that also passed through the target aperture.

After selecting the data to be fit, we used the data in the background
annulus to measure and subtract an initial background.  We then fitted
a two dimensional elliptical Gaussian function to the data in both the
target aperture and the background annulus.  The seven free parameters
in this fit were the peak voltage (the difference between the on-
  and off-source voltages, which is a negative value because an
increase in signal corresponds to a decrease in voltage), the right
ascension and declination of the peak position, the major and minor
axes of the beam, the position angle of the axes, and the background.

\begin{figure}
\begin{center}
\epsfig{file=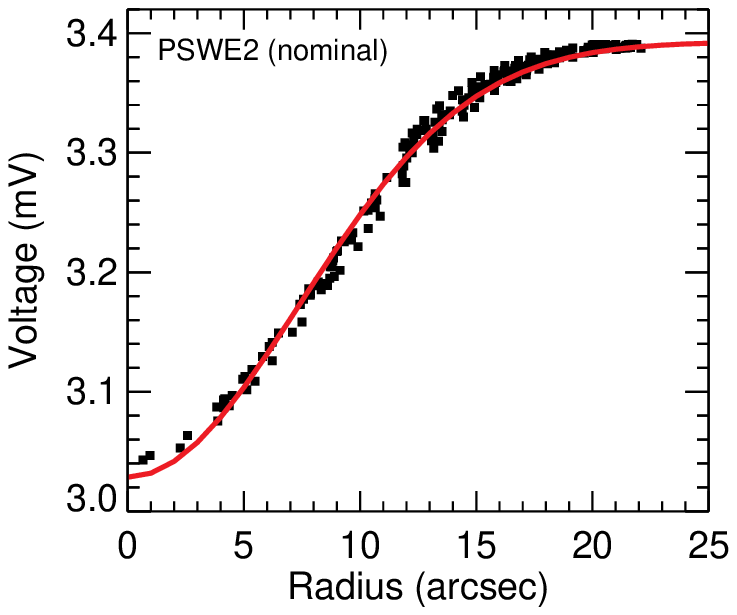}
\epsfig{file=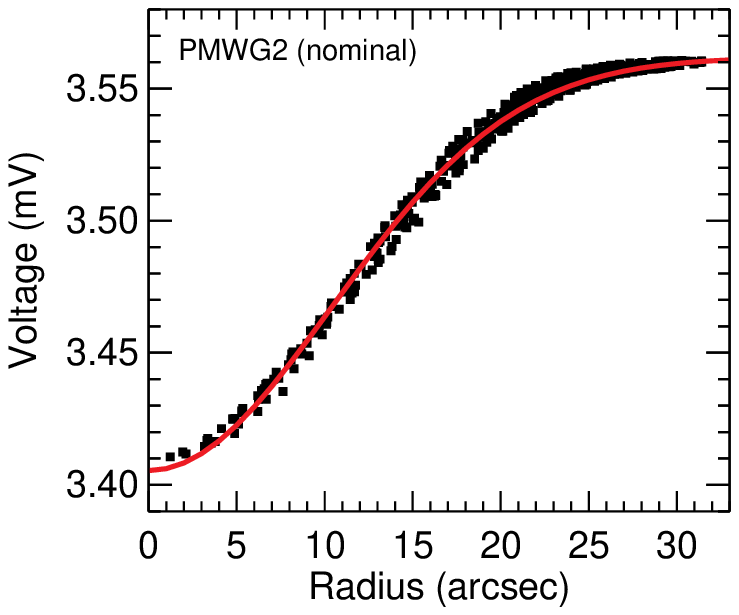}
\epsfig{file=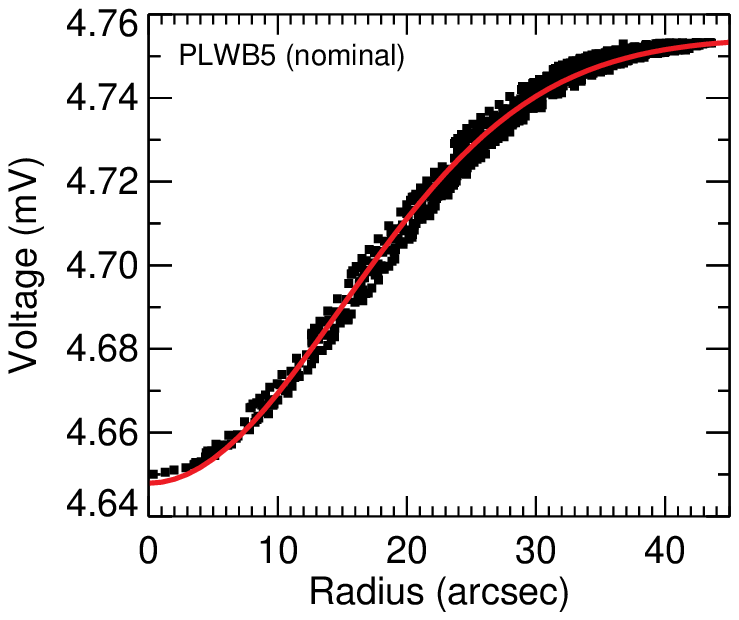}
\caption{Examples of the signal measured from Neptune in the fine scan
observation 1342186522, which used the nominal voltage bias setting.
The red line shows the best fitting function to the data.  Data
falling within the background annulus were included in the fit but are
not included in this plot.}
\label{f_psffit_nominal}
\end{center}
\end{figure}

\begin{figure}
\begin{center}
\epsfig{file=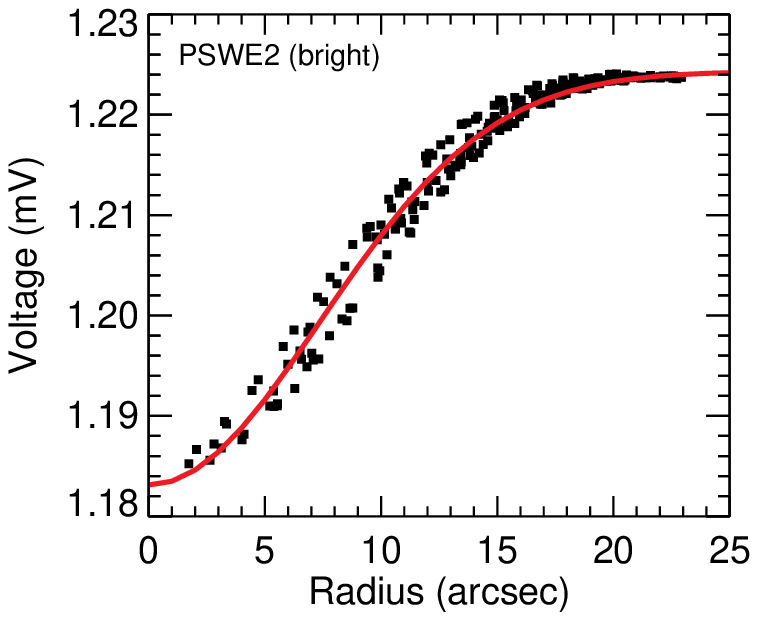}
\epsfig{file=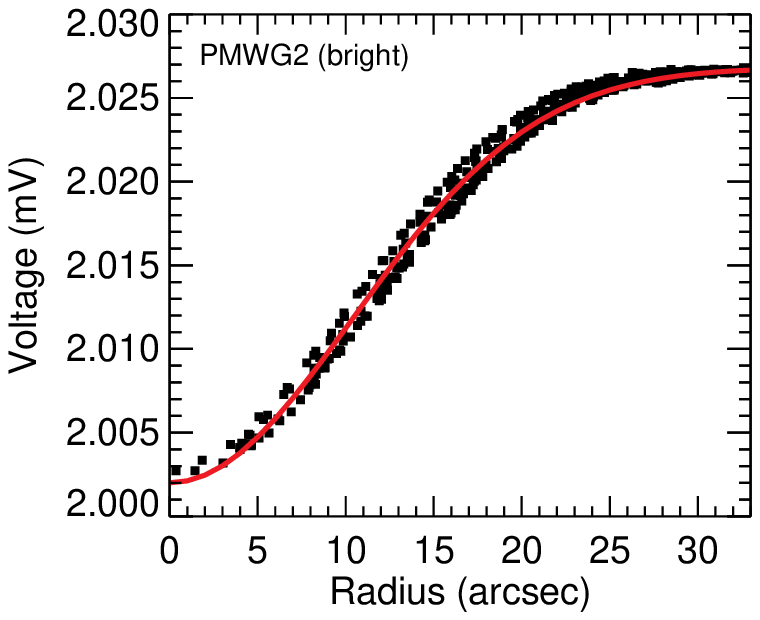}
\epsfig{file=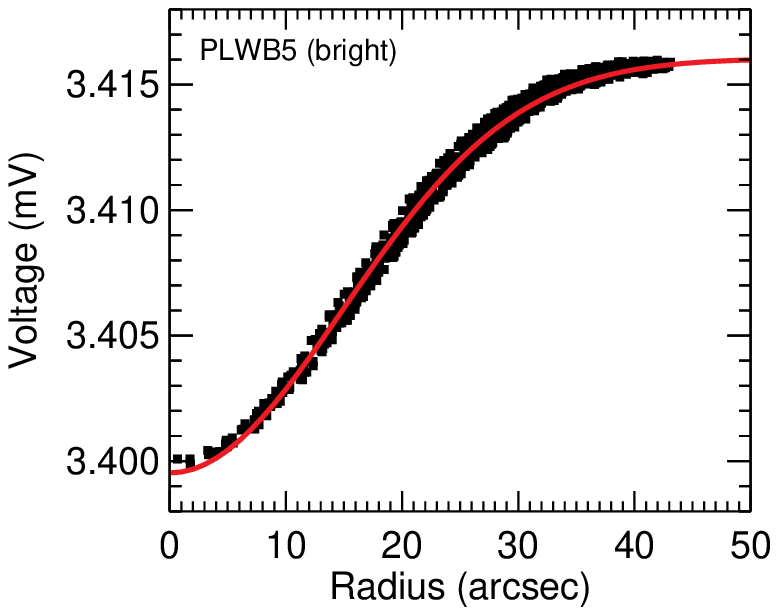}
\caption{Examples of the signal measured from Neptune in the fine scan
observation 1342187438, which used the bright source bias setting.
The red line shows the best fitting function to the data.  Data
falling within the background annulus were included in the fit but are
not included in this plot.}
\label{f_psffit_bright}
\end{center}
\end{figure}

Example radial profiles of the data from within the target apertures
and fits to those data for the nominal and bright source mode data are
shown in Figures~\ref{f_psffit_nominal} and \ref{f_psffit_bright}.
Although the beam is well fitted by Gaussian functions in these
examples, the data may deviate from the fit at the $\sim1$\% level
near the peak of the beam.  We also explored using alternate functions
(e.g. sinc$^2$ functions, polynomial functions) to determine the peak
of the beam, but we found that the peak signal varied by $\sim1$\%
when slight alterations were made to the fitting method (such as
changing the radius within which data were selected and adjusting the
exact shape of the fitted function).  Given the simplicity of using
Gaussian functions and the lack of any advantages (in particular, the
lack of any improvement in accuracy) in using other functions to fit
the data, we based the calibration on fitting Gaussian functions.

\begin{figure}
\epsfig{file=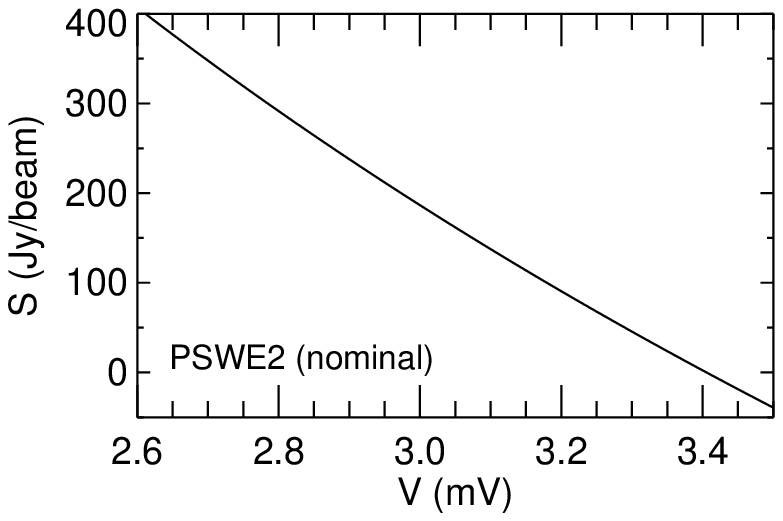}
\epsfig{file=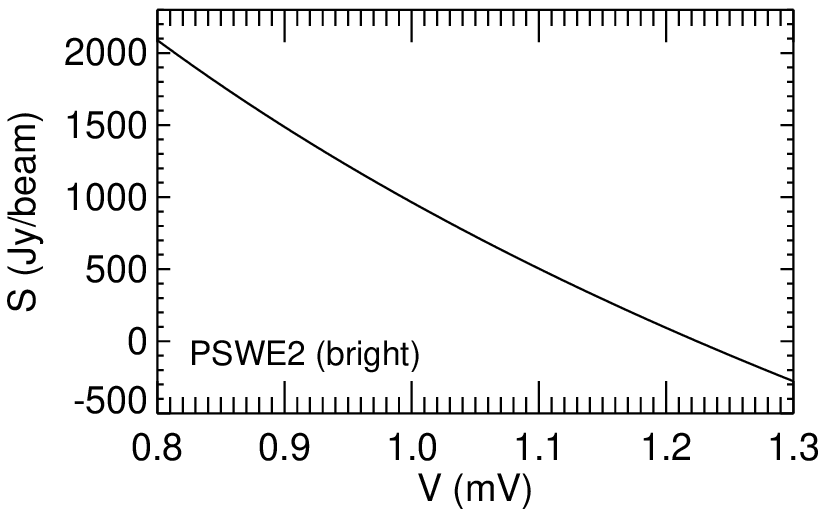}
\caption{Calibration curves for PSWE2.  Data are shown for both bias modes.}
\label{f_calcurve}
\end{figure}

For each bolometer in each bias voltage setting, we generally obtained
a set of four Neptune peak voltage measurements with corresponding
background voltage measurements.  We used the unscaled $K$-parameters
from the PCal flash analysis in Section~\ref{s_curve_derivation} in
Equation~\ref{e_cal} to determine the change in voltage between the
background and Neptune for each bolometer for each observation.  To
produce the scaling terms $A$, we then divided the measured change in
signal by the Neptune model flux densities (with beam corrections
applied to account for the finite angular size of Neptune) given in
Table~\ref{t_nepflux} and by the $K_{MonP}$ values given in
Table~\ref{t_kmonp}.  For each bolometer in each voltage mode, the
unscaled $K_1$ and $K_2$ parameters are divided by the mean of the $A$
terms from the four observations, and the standard deviation in the
four $A$ terms is used to calculate the uncertainties in the $K_1$ and
$K_2$ parameters.  Figure~\ref{f_calcurve} shows examples of the
resulting scaled calibration curves from the nominal and bright source
modes.  See Section~\ref{s_unc_nep} for the analysis on the fractional
uncertainties related to the scaling terms.

\begin{table*}
\centering
\begin{minipage}{175mm}
\caption{Neptune flux densities on dates of fine scan observations}
\label{t_nepflux}
\begin{tabular}{@{}cccccccccccc@{}}
\hline
Operation &    
    Observations &    
    Voltage &    
    \multicolumn{3}{c}{SRF-weighted Flux Densities} &
    \multicolumn{3}{c}{Beam Correction Factors$^a$} &
    \multicolumn{3}{c}{SRF-weighted Flux Densities}\\
Day &    
    &    
    Bias Mode &    
    \multicolumn{3}{c}{without Beam Correction (Jy)} &
    \multicolumn{3}{c}{} &
    \multicolumn{3}{c}{with Beam Correction (Jy)}\\
&          
    &                 
    &
    250~$\mu$m &    350~$\mu$m &    500~$\mu$m &
    250~$\mu$m &    350~$\mu$m &    500~$\mu$m &
    250~$\mu$m &    350~$\mu$m &    500~$\mu$m \\
\hline
168 (29 Oct 2009) &
    1342186522 &
    nominal &
    163.48 &        102.85 &        61.26 &
    0.99419 &       0.99684 &       0.99853 &
    162.53 &        102.52 &        61.17 \\
&   1342186523 & & & & & & & & & & \\
&   1342186524 & & & & & & & & & & \\
&   1342186525 & & & & & & & & & & \\
201 (01 Dec 2009) &
    1342187507 &
    bright &
    157.45 &        99.06 &         59.00 &
    0.99440 &       0.99696 &       0.99859 &
    156.57 &        98.75 &         58.92 \\
&   1342187508 & & & & \\
202 (02 Dec 2009) &
    1342187438 &
    bright &
    157.28 &        98.95 &         58.94 &
    0.99441 &       0.99696 &       0.99859 &
    156.40 &        98.65 &         58.85 \\
&   1342187439 & & & & \\
\hline
\end{tabular}
$^a$ The beam correction factors account for the finite angular size of Neptune
and are based on a Gaussian main beam coupling to the emission from 
planetary disc \citep{uh76}.
\end{minipage}
\end{table*}

\subsubsection{Truncated detector signals in Neptune fine scan data}
\label{s_scale_trunc}

In the nominal voltage bias data, 17 detectors in the PSW array and 6
detectors in the PMW array measured signals from Neptune that were
truncated in at least one of the observations.  In the SPIRE on-board
electronics, an individual voltage offset is applied to each bolometer
signal to ensure that its output voltage remains within the range of
the Analogue to Digital Converter (ADC). The offsets are fixed at the
start of each observation and depend on the exact telescope background
and sky brightness. Neptune is sufficiently bright that its large
signal level drove a few bolometer voltage levels outside the ADC
range, leading to signal truncation.  In the analysis of these fine
scan data, we needed to alter our analysis for these truncated
bolometers.

We examined the effects of the signal truncation by experimenting with
how the peak signal measured by the ``good'' bolometers changed as we
progressively removed data from the timelines (with ``good''
bolometers defined as those where the Neptune timelines are not
truncated in any of the nominal-voltage fine scan observations and
where the bolometers are not identified as problematic as described in
Section~\ref{s_probbolom}).  For each ``good'' bolometer in each
nominal voltage observation, we first fitted all of the data, then
artificially truncated the signal by removing data above a set
fraction $R_{actual}$ of the best fitting peak voltage value, and then
we repeated the fit. The functions fitted to the data were elliptical
Gaussian function where the axes and position angle were fixed to the
mean values measured for the individual bolometers in the bright
source mode timeline data\footnote[16]{To assess the appropriateness
  of using bright source mode beam dimensions to represent nominal
  mode beams in both the PSW and PMW data, we compared the bright
  source and nominal voltage mode measurements of the major axis,
  minor axis, and position angle for each non-problematic bolometer
  where the signal from Neptune was not truncated.  The major and
  minor axes typically vary by \ltsim2 \% between the modes, and the
  position angles of the axes only vary by $\ltsim10^\circ$.}.  We
found that fixing the dimensions of the beam in this analysis produced
the results with the lowest variance.  We then calculated the mean and
standard deviation in the ratio of the best fitting peak voltage for
the truncated data compared to the value measured when all data were
used.  These ratios are the correction factors for the peak fits to
bolometers that saturate on Neptune.  As the fitted peak value may
change when the data are removed from the peak, we also calculated the
ratio of the lowest voltage values (which correspond to the highest
signal values) in the fitted data to the peak voltage derived from the
fit to those data (which we label as $R_{meas}$).  Also note that the
results of this analysis apply specifically to the elliptical Gaussian
functions fit to the beams and may not necessarily be applicable when
other functions are fit to the data.

Table~\ref{t_satbolomcorr} lists the correction factors derived using
the above analyses.  The peak voltage measurements from the truncated
data were divided by correction factors that were found by
interpolating between the values listed in Table~\ref{t_satbolomcorr}.
Table~\ref{t_satbolomdata} lists the mean $R_{meas}$ values measured
in the fine scan observations for each bolometer where the Neptune
signal is truncated as well as the mean correction that is applied for
each bolometer.  Because offset settings sometimes changed between
fine scan observations, not all of the Neptune data from the
bolometers listed in Table~\ref{t_satbolomdata} were truncated in all
of the observations.  These specific cases have been flagged in the
table.  Additionally, data from PMWB4, PMWC7, and PMWB8 were each
completely truncated in one of the fine scan observations.  In these
cases, we ignored data for any bolometer from any observation in which
all of the data were truncated, applied corrections when only the peak
was truncated, and used the direct peak measurements when the data are
not truncated.

\begin{table}
\caption{Correction factors for peak voltage$^a$ measurements for
   bolometers where the Neptune fine scan data is truncated}
\label{t_satbolomcorr}
\begin{center}
\begin{tabular}{@{}lccc@{}}
\hline
Array &    $R_{actual}~^{b}$ &     Correction$^{c}$&         $R_{meas}~^{d}$\\
\hline
PSW &      0.900 &              $1.012 \pm 0.005$ &      0.890\\
    &      0.800 &              $1.017 \pm 0.007$ &      0.786\\
    &      0.700 &              $1.023 \pm 0.009$ &      0.684\\
    &      0.600 &              $1.025 \pm 0.011$ &      0.586\\
    &      0.500 &              $1.018 \pm 0.014$ &      0.491\\
    &      0.400 &              $1.005 \pm 0.017$ &      0.398\\
    &      0.300 &              $0.977 \pm 0.021$ &      0.307\\
    &      0.200 &              $0.929 \pm 0.030$ &      0.215\\
PMW &      0.900 &              $1.008 \pm 0.007$ &      0.893\\
    &      0.800 &              $1.012 \pm 0.011$ &      0.790\\
    &      0.700 &              $1.014 \pm 0.018$ &      0.690\\
    &      0.600 &              $1.012 \pm 0.020$ &      0.593\\
    &      0.500 &              $1.006 \pm 0.024$ &      0.497\\
    &      0.400 &              $0.991 \pm 0.029$ &      0.404\\
    &      0.300 &              $0.963 \pm 0.033$ &      0.312\\
\hline
\end{tabular}
\end{center}
$^a$ The peak voltage is defined as the difference between the on-target
     and off-target voltages.\\
$^{b}$ In the ``good'' bolometer data used to derive the corrections in
     this table, $R_{actual}$ is the ratio of the lowest voltage values
     used for fitting the data to the best fitting peak
     voltage (when all data were used).\\
$^{c}$ The best fitting peak voltage measurement from the truncated signal
     data should be divided by these correction factors.\\
$^{d}$ $R_{meas}$ represents the ratio between the lowest voltage values used
     for fitting the data to the peak voltage determined from fitting those
     data.
\end{table}

\begin{table}
\caption{Bolometers with truncated Neptune signal data in nominal voltage mode}
\label{t_satbolomdata}
\begin{center}
\begin{tabular}{@{}lcc@{}}
\hline
Bolometer &    Mean &            Mean \\
          &    $R_{meas}~^a$ &   Correction$^b$\\
\hline
PSWA4     &    0.200 &           0.919 \\
PSWB5     &    0.544 &           1.023 \\
PSWB11    &    0.844 &           1.014 \\
PSWC4     &    0.787 &           1.017 \\
PSWC6     &    0.621 &           1.025 \\
PSWD2     &    0.394 &           1.004 \\
PSWD4     &    0.853 &           1.014 \\
PSWD12    &    0.641 &           1.025 \\
PSWE6     &    0.815 &           1.015 \\
PSWE10    &    0.435 &           1.011 \\
PSWE14    &    0.298 &           0.973 \\
PSWF4$^c$ &    0.144 &           0.879 \\
PSWF12    &    0.323 &           0.983 \\
PSWG1     &    0.254 &           0.952 \\
PSWG3     &    0.762 &           1.018 \\
PSWH2     &    0.358 &           0.995 \\
PSWJ14    &    0.552 &           1.023 \\
PMWB1$^c$ &    0.385 &           0.986 \\
PMWB8$^c$ &    0.368 &           0.982 \\
PMWC2$^c$ &    0.386 &           0.987 \\
PMWD3     &    0.756 &           1.013 \\
PMWD12    &    0.754 &           1.013 \\
PMWF3     &    0.728 &           1.014 \\
\hline
\end{tabular}
\end{center}
$^a$ $R_{meas}$ represents the ratio between the lowest voltage values used
     for fitting the data to the peak voltage determined from fitting those
     data.\\
$^b$ The best fitting peak voltage measurement from the truncated signal
     data should be divided by these correction factors.\\
$^c$ For these bolometers, the signal from Neptune is truncated in
     some but not all of the observations.  The numbers here are
     for only the observations in which the peak signal from Neptune
     is truncated but the background signal is not truncated.\\
\end{table}

\section{Problematic bolometers}
\label{s_probbolom}

For some bolometers that have been identified as dead, noisy, or slow,
it was still possible to measure $V$ and $\Delta V$ values from the
PCal flash data and the amplitude of the voltage from the Neptune fine
scan data, which allows us to derive calibration curves.  However, for
others, we lacked usable measurements from either the PCal flash data
or the fine scan data, and so we cannot create empirical calibration
curves for these bolometers.  For these bolometers, we needed to
insert alternate $K$ values as placeholders into the flux calibration
table to avoid problems when executing the flux calibration software.
We used the values derived from the bolometer models that are listed
in version 2-3 of the flux calibration table, which are derived from models
of the bolometer responsivities readjusted based on early observations
of Ceres and Alpha Boo.  Table~\ref{t_dnsbolom} lists the bolometers
that are labelled as dead, noisy, or slow as well as an indication of
whether empirical calibration curves could be derived for the
bolometers.  Separate attempts were made to calibrate bolometers in
the nominal and bright source voltage modes, but we obtained identical
results in terms of being able to derive calibration curves.

\begin{table}
\caption{Calibration status of dead, noisy, or slow bolometers}
\label{t_dnsbolom}
\begin{center}
\begin{tabular}{@{}lcc@{}}
\hline
Bolometer &  Problem &  Empirical Calibration$^a$ \\
\hline
PSWA10 &     Slow  &    Y \\
PSWA11 &     Slow  &    N \\
PSWA13 &     Slow  &    N \\
PSWC12 &     Dead  &    N \\
PSWD15 &     Dead  &    N \\
PSWF9  &     Noisy &    Y \\
PSWG8  &     Dead  &    N \\
PSWG11 &     Dead  &    N \\
PMWA13 &     Slow  &    N \\
PMWB11 &     Noisy &    Y \\
PMWD1  &     Dead  &    Y \\
PMWD6  &     Noisy &    Y \\
PMWE8  &     Noisy &    Y \\
PLWA6  &     Dead  &    N \\
PLWC9  &     Noisy &    Y  \\
\hline
\end{tabular}
\end{center}
$^a$ An entry of ``Y'' indicates that an empirical calibration was
derived using the PCal flash and Neptune fine scan data.  An entry of ``N''
indicates that this was not possible.\\
\end{table}

All of the subsequent tests that we performed with the new flux
calibration values did not include any of the bolometers in
Table~\ref{t_dnsbolom}.  They are currently flagged as bad bolometers
in the timeline data and are not used by default in the mapmaking, and
so it would be inappropriate to include them in the tests.

\section{Calibration uncertainty budget for individual bolometers}
\label{s_unc}

The conversion between detector signal and astronomical signal for
individual bolometers in the SPIRE photometer arrays involves four
sources of uncertainty.  Two of these sources of uncertainty are
systematic effects across all bolometers in each array and are
independent of the measurements used to derive the $K$-parameters.
These systematic effects are the absolute uncertainty in the Neptune
model flux density and the uncertainty related to uncertainties in the
SPIRE SRFs.  The uncertainty in the Neptune model flux density is 4\%.
The uncertainty related to the SPIRE SRFs is dominated by
uncertainties in the positions of the band edges; uncertainties in the
shapes of the SRFs have less of an effect.  Griffin et al. (2013,
submitted) determined that the uncertainty in the scaling of the
individual bolometers resulting from uncertainties in the SRF peaks at
1.6\%.

The other two sources of uncertainty in the the flux calibration of
the individual bolometers are associated with the measurements used to
derive the $K$-parameters: uncertainties from fitting the
PCal flash data (Section~\ref{s_curve_derivation}) and uncertainties
from determining the scaling terms (Section~\ref{s_scale}).  We
quantify these uncertainties below.

\subsection{Uncertainty from fitting the PCal flash data}

In theory, it should be possible to derive uncertainties in the
unscaled versions of the $K$-parameters from the fits to the data.
However, degeneracy problems arose when attempting to fit
Equation~\ref{e_fdefinition} to the PCal flash data; small variations
in the input data could result in significant variations in the
$K$-parameters, including significant variations in the relative
magnitudes of the first and second terms on the right side of
Equation~\ref{e_fdefinition}.  We therefore could not directly
calculate uncertainties for these terms.

Instead, we calculated how the uncertainties in the PCal flash
measurements would affect the final calibration curve using a Monte
Carlo technique.  In a series of 1000 trials, random noise scaled by
the uncertainties in the magnitude of the PCal flash measurements was
added to the original PCal flash data.  Equation~\ref{e_fdefinition}
was then fitted to the data, and the $K$-parameters were rescaled
using the $A$ parameters from the analysis in Section~\ref{s_scale}.
The standard deviation in all of these calibration curves was measured
at evenly spaced locations in voltage.  This standard deviation curve
was then used as the uncertainty in the flux density $\bar{S}_{Meas}$
resulting from the uncertainties in the PCal flash measurements.

\begin{figure}
\epsfig{file=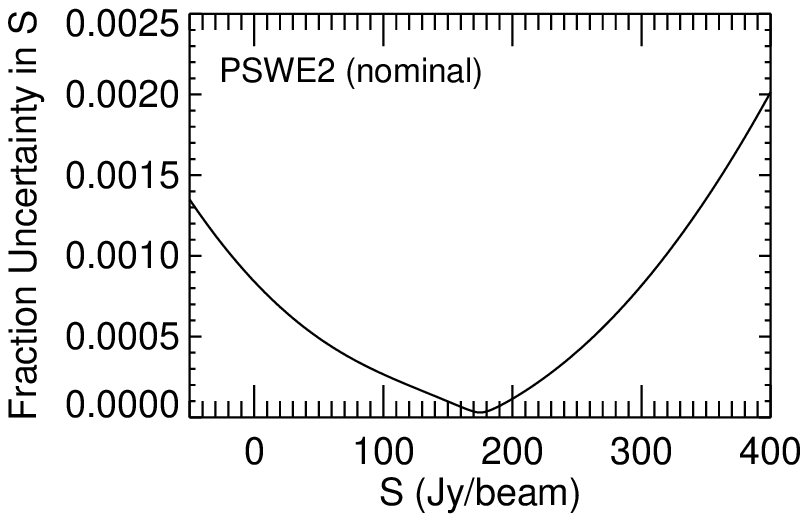}
\epsfig{file=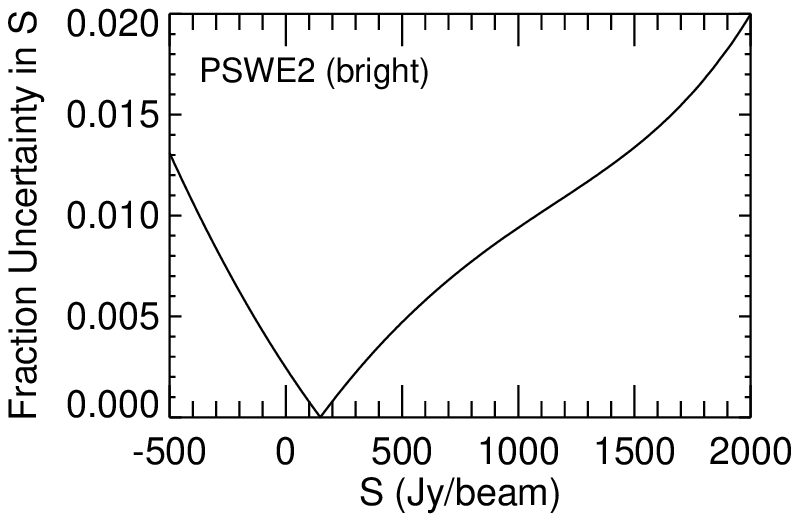}
\caption{Uncertainties in the calibration curves related to the fits
  to the PCal flash data for PSWE2. Curves are shown for both bias
  modes.}
\label{f_fracunccurve}
\end{figure}

Examples of the resulting fractional uncertainty curves are shown in
Figure~\ref{f_fracunccurve}.  Because Neptune was used to derive the
scaling terms, the uncertainty curves drop to $\sim0$ at locations
that correspond to the flux densities of Neptune.
Table~\ref{t_fracunccurve} states the median fractional uncertainties
measured for the curves within the regions covered by the PCal flash
data.

\begin{table}
\caption{Median fractional uncertainties related to fits from PCal flash data}
\label{t_fracunccurve}
\begin{center}
\begin{tabular}{@{}lcc@{}}
\hline
Array &      \multicolumn{2}{c}{Median Fractional Uncertainty}\\
&            Nominal &      Bright Source\\
&            Mode &         Mode\\
\hline
PSW &        0.00021 &      0.0060 \\
PMW &        0.00022 &      0.0052 \\
PLW &        0.00017 &      0.0020 \\
\hline
\end{tabular}
\end{center}
\end{table}

For the nominal voltage mode, the median fractional uncertainties from
the fits to the PCal flash data are very small compared to the
uncertainties related to the variance in the peak flux measurements of
Neptune (see Section~\ref{s_unc_nep}).  This is mainly because the
large amount of PCal flash data that is available has tightly
constrained the calibration curves.  The uncertainties for the bright
source mode data are higher because fewer observations with PCal flashes were
performed, and these median fractional uncertainties are comparable to
the median fractional uncertainties in the scaling terms based on the
Neptune data.

\subsection{Uncertainty in the scaling terms for the calibration curves}
\label{s_unc_nep}

As described in Section~\ref{s_scale}, the fine scan observations
produced a total of four peak voltage measurements of Neptune for each
bolometer.  The standard deviation in the $A$ values from these four
measurements was used to derive the uncertainty in the flux
calibration for each detector.  Tables~\ref{t_fracuncnep_nominal} and
\ref{t_fracuncnep_bright} give the mean and maximum uncertainty in the
scaling terms based on the results for all bolometers in each array
and in each bias voltage mode.

\begin{table*}
\centering
\begin{minipage}{115mm}
\caption{Uncertainties related to variance in peak flux measurements
    of Neptune for the nominal mode}
\label{t_fracuncnep_nominal}
\begin{tabular}{@{}lcccccc@{}}
\hline
Array &      \multicolumn{6}{c}{Fractional Uncertainties} \\
&            \multicolumn{2}{c}{(all non-problematic} &
             \multicolumn{2}{c}{(non-problematic bolometers } &
             \multicolumn{2}{c}{(non-problematic bolometers } \\
&            \multicolumn{2}{c}{bolometers)} &
             \multicolumn{2}{c}{with truncated Neptune data)} &
             \multicolumn{2}{c}{with no truncated Neptune data)} \\
&            Median &              Maximum &
             Median &              Maximum &
             Median &              Maximum \\
\hline
PSW &        0.0059 &              0.047 &
             0.022 &               0.047 &
             0.0049 &              0.023 \\
PMW &        0.0042 &              0.045 &
             0.016 &               0.045 &
             0.0039 &              0.019 \\
PLW &        0.0052 &              0.012 &   
             $^a$ &                $^a$ &
             0.0052 &              0.012 \\
\hline
\end{tabular}
$^a$ None of the PLW bolometers were truncated during the Neptune fine scan
observations.
\end{minipage}
\end{table*}

The nominal mode fractional uncertainties are reported for all
non-problematic bolometers (bolometers not identified as dead, noisy,
or slow in Table~\ref{t_dnsbolom}) and then for subsets of the
non-problematic bolometers where the signal from Neptune in the fine
scan data was either truncated (listed in Table~\ref{t_satbolomdata})
or not truncated.  The fractional uncertainties for the bolometers
with untruncated Neptune data typically stayed below 1\%.  The
uncertainties for the bolometers with truncated Neptune data, however,
are typically $\sim2$\% and may approach 5\%, reflecting the inherent
uncertainties in the correction factors for the truncation.  The worst
individual bolometers were PSWF4, which had high uncertainties because
the Neptune data were severely trucated in two of the fine scan
observations, and PMWB8, which had high uncertainties partly because
the Neptune data were strongly truncated and partly because one of the
fine scan observations produced no usable data for the bolometer.  The
uncertainties reported in Table~\ref{t_fracuncnep_nominal} are
generally much higher than the uncertainties from the fits to the PCal
flash data listed in Table~\ref{t_fracunccurve}, which means that
these uncertainties in scaling the calibration curves dominate the
overall instrumental flux calibration uncertainties for individual
bolometers.

For the bright source bias setting, the median uncertainties in the
Neptune measurements are $<1$\%.  For the 250 and 350~$\mu$m arrays,
the uncertainties from fitting Equation~\ref{e_fdefinition} to the
PCal flash data is the dominant instrumental source of uncertainty for
individual bolometers, as can be seen by comparing the median
fractional uncertainties in Tables~\ref{t_fracuncnep_bright} and
\ref{t_fracunccurve}.  However, the uncertainties in the scaling terms
measured for the 500~$\mu$m bolometers are greater than the
uncertainties from fitting the PCal flash data.

\section{Tests of the flux calibration}
\label{s_test}

While the assessments in Section~\ref{s_unc} quantify the
uncertainties in the calibration of individual bolometers, they are
not indicative of the flux calibration uncertainties applicable to a
source observed by multiple bolometers in standard scan map
observations.  We tested the point source flux calibration scheme
using scan map observational data of Neptune, Uranus, and the star
Gamma Dra.  Uranus was used for these tests because its flux density
is known as accurately as Neptune's.  Gamma Dra is a K5{\small III}
star \citep{pe97} used as a SPIRE secondary calibrator that is
observed regularly and that is visible at all times of the year.
While the flux density of the star is not known as accurately as the
flux densities of Neptune and Uranus, the large amount of data
acquired for Gamma Dra during the course of the mission makes it
particularly useful for examining the consistency and long-term
stability of the flux calibration.  None the less, the tests with
Neptune are the most important mainly because they demonstrate the
repeatability of flux density measurements of the primary calibration
source.

We used the standard SPIRE scan map data processing pipeline in HIPE
version 10.0.620 and the default calibration products from version
8\_1 of the calibration tree.  We removed the temperature drift
removal from the pipeline because it is dependent on the flux
calibration, but to compensate for this, we used the baseline removal
module to remove the temperature drift and other variations in the
background.  The data used for these tests were standard scan map
(large scan map), standard small scan map, and standard parallel
observing mode data for the sources.  Almost all small and large scan
map observations were performed using the medium or nominal scan speed
(30 arcsec s$^{-1}$), although one or two large scan map observation
of each target were performed using the fast scan speed (60 arcsec
s$^{-1}$), two of the Gamma Dra parallel mode observations used the
fast scan speed, and the other two Gamma Dra parallel mode
observations used the slow scan speed (20 arcsec s$^{-1}$).  All
available observations of these sources that were performed between OD
100 (the date of the first calibration observations after the bias
voltages were set to their current levels) and OD 1434 (the last date
in the mission when SPIRE photometer observations of flux calibration
sources were performed) were used.  Flux densities were measured by
performing fits to the timeline data using a similar method described
in Section~\ref{s_scale}, but we used all unflagged data from all
bolometers in each array instead of the data from single bolometers.

\begin{table}
\caption{Uncertainties related to variance in peak flux measurements
    of Neptune for the bright source mode}
\label{t_fracuncnep_bright}
\begin{center}
\begin{tabular}{@{}lcc@{}}
\hline
Array &      \multicolumn{2}{c}{Fractional Uncertainties$^a$} \\
&            Median &              Maximum \\
\hline
PSW &        0.0032 &              0.0073 \\
PMW &        0.0023 &              0.012 \\
PLW &        0.0038 &              0.014 \\
\hline
\end{tabular}
\end{center}
$^a$ The values reported here exclude the uncertainties for the dead,
noisy, and slow bolometers listed in Table~\ref{t_dnsbolom}.
\end{table}

\begin{figure*}
\begin{center}
\epsfig{file=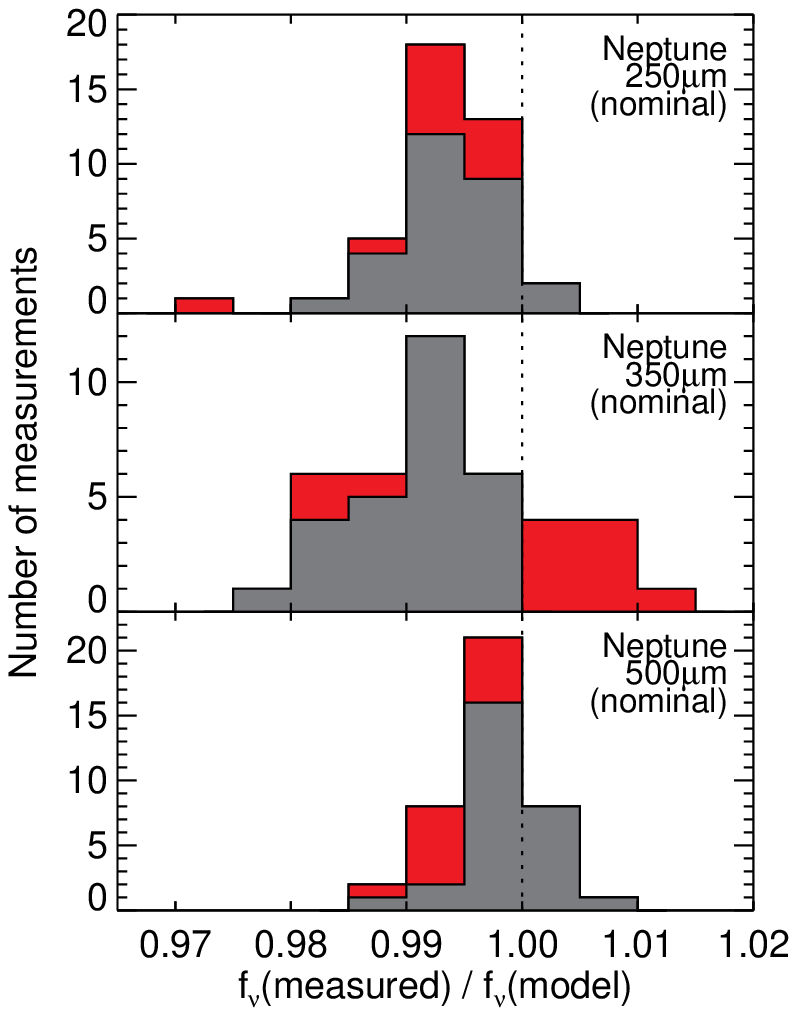}
\epsfig{file=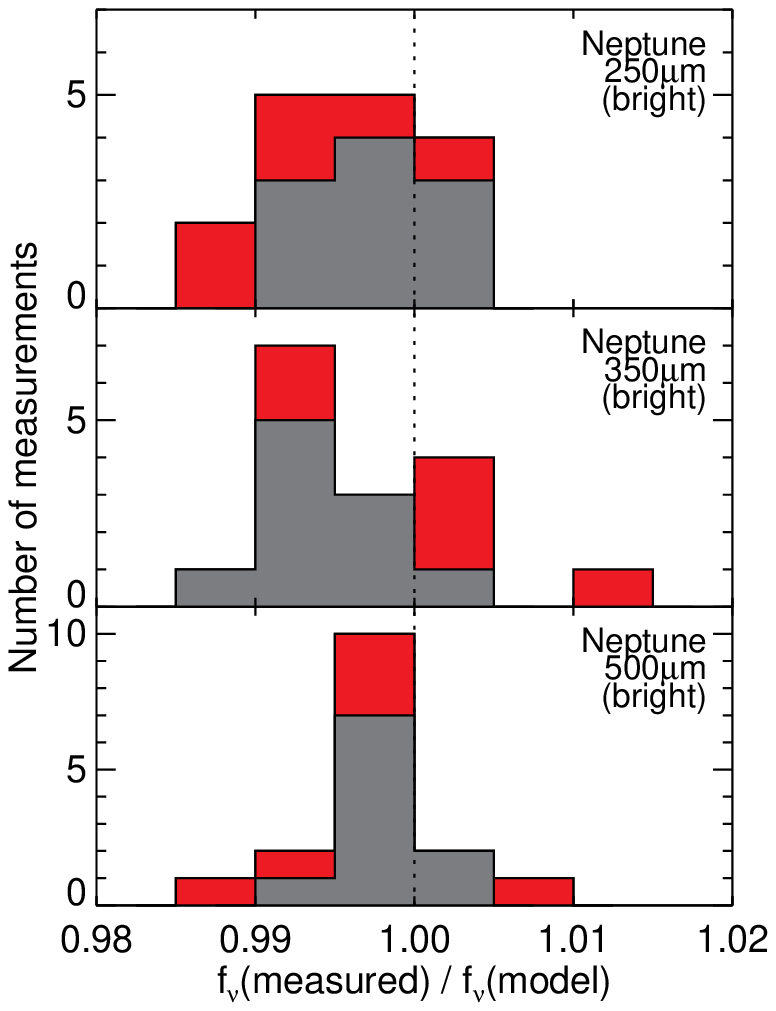}
\end{center}
\caption{Histograms of the measured to model flux densities for
  Neptune.  The grey data represent measurements made in small scan map
  mode, and the red data represent measurements made in large scan map
  mode.}
\label{f_hist_nep}
\end{figure*}

\begin{figure*}
\begin{center}
\epsfig{file=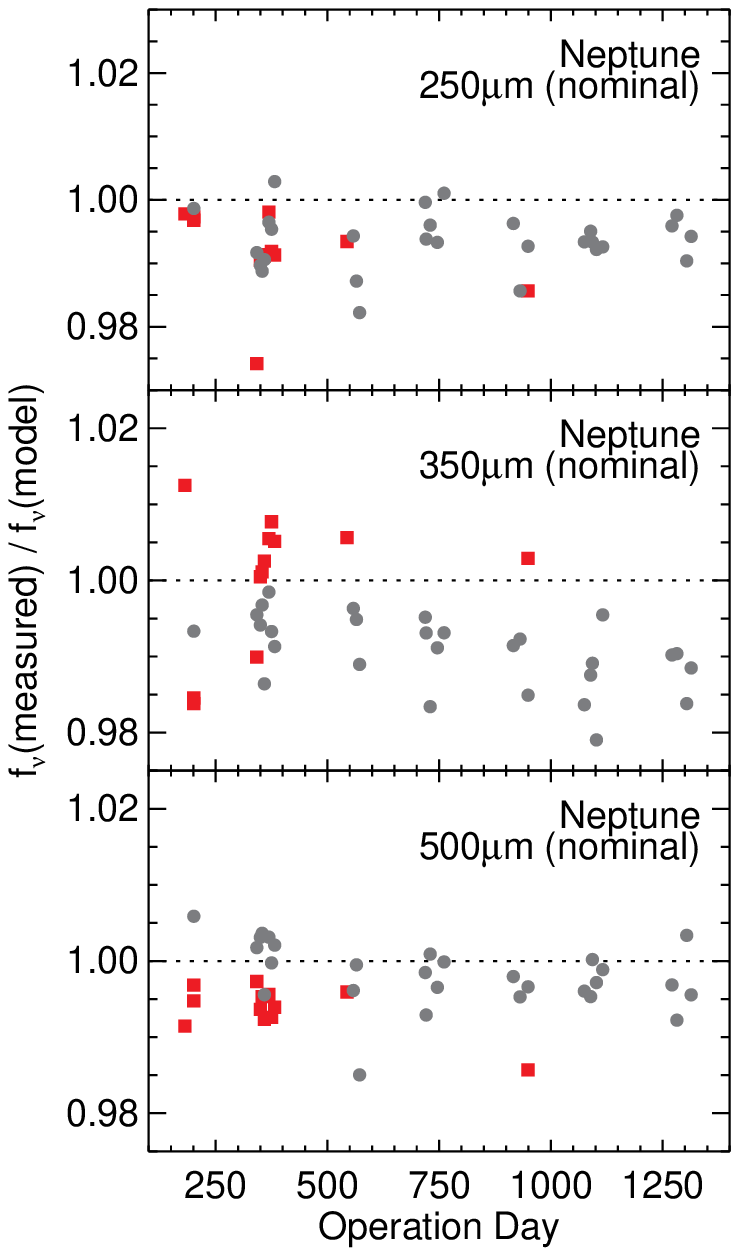}
\epsfig{file=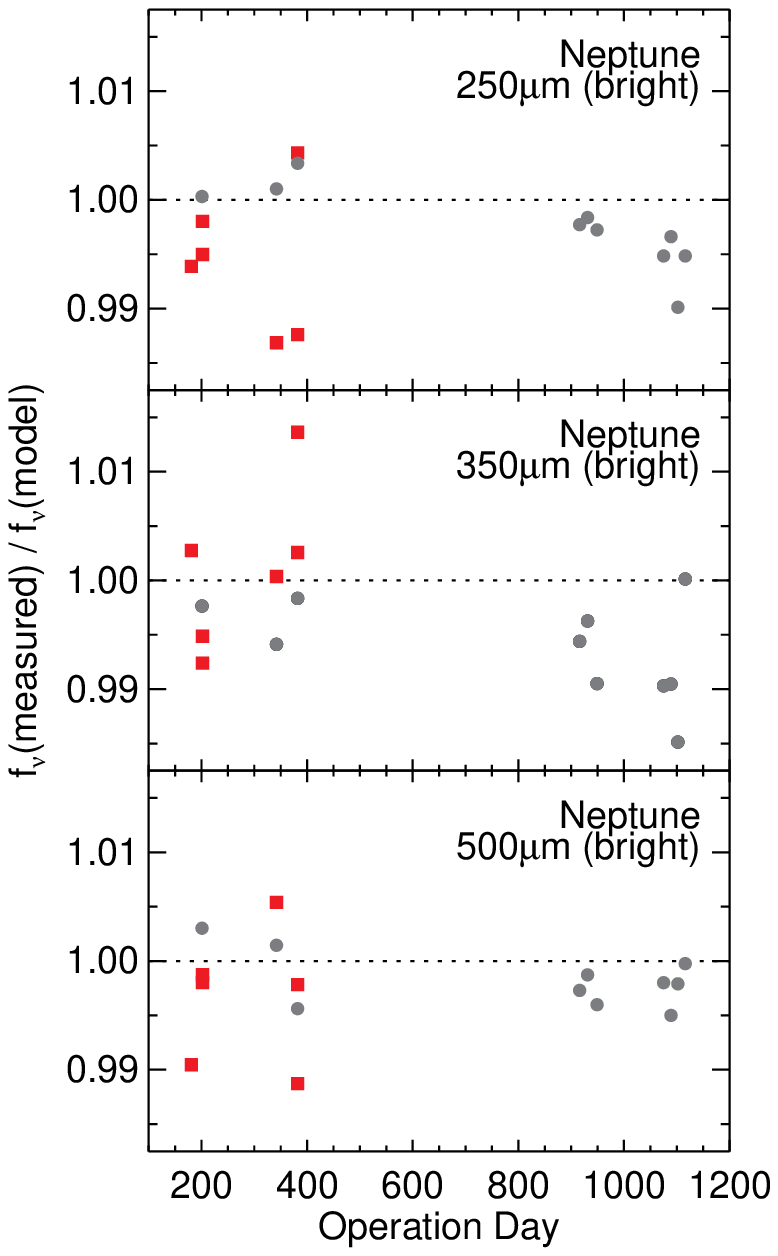}
\end{center}
\caption{Ratios of the measured to model flux densities for Neptune
plotted as a function of OD.  The grey circles are measurements made
in small scan map data, and the red squares are measurements made in
large scan map data.  The uncertainties from the fits are equivalent
to or smaller than the symbols in these plots.}
\label{f_test_nep}
\end{figure*}

\subsection{Tests with Neptune data}

The flux density of Neptune varies over time because of changes in the
distance between it and the Earth, so we cannot simply report
statistical results on the flux densities.  Instead, we divided the
measured flux densities by the model flux densities so that data from
different ODs could be compared together.  These ratios are shown in
Figure~\ref{f_hist_nep} and Table~\ref{t_test_nep}, and
Figure~\ref{f_test_nep} shows how the ratios vary over time.

\begin{table*}
\centering
\begin{minipage}{141mm}
\caption{Measured/model flux density ratios for Neptune}
\label{t_test_nep}
\begin{tabular}{@{}lcccccccc@{}}
\hline
Waveband &   \multicolumn{6}{c}{Measured/Model Flux Density Ratios} &
             \multicolumn{2}{c}{Change in Ratios between}\\
($\mu$m)&    \multicolumn{3}{c}{Nominal Mode} &
             \multicolumn{3}{c}{Bright Source Mode} &
             \multicolumn{2}{c}{OD 100 and OD 1450$^a$}\\
&            All &             Small &                Large &
             All &             Small &                Large &
             Nominal &         Bright\\
&            Maps &            Scan Maps &            Scan Maps &
             Maps &            Scan Maps &            Scan Maps &
             Mode &            Source Mode\\
\hline
250 &        $0.993$ &         $0.994$ &              $0.992$ &
             $0.997$ &         $0.997$ &              $0.994$ &
             $0.000$ &
             $-0.011$\\    
    &        $\pm0.005$ &      $\pm0.005$ &           $\pm0.007$ &
             $\pm0.005$ &      $\pm0.004$ &           $\pm0.007$ &
             $\pm0.003$ &
             $\pm0.003$\\    
350 &        $0.993$ &         $0.991$ &              $1.003$ &
             $0.996$ &         $0.994$ &              $1.001$ &
             $-0.010$ &
             $-0.008$\\
    &        $\pm0.008$ &      $\pm0.005$ &           $\pm0.009$ &
             $\pm0.007$ &      $\pm0.005$ &           $\pm0.007$ &
             $\pm0.003$ &
             $\pm0.005$\\
500 &        $0.997$ &         $0.998$ &              $0.994$ &
             $0.998$ &         $0.998$ &              $0.998$ &
             $-0.006$ &
             $-0.005$\\
    &        $\pm0.004$ &      $\pm0.004$ &           $\pm0.003$ &
             $\pm0.004$ &      $\pm0.003$ &           $\pm0.006$ &
             $\pm0.003$ &
             $\pm0.003$\\
\hline
\end{tabular}
$^a$ These values are calculated by fitting lines to the
    measured/model ratios for the small scan map data and then
    calculating the difference in the best fitting line between OD 100
    and OD 1450.
\end{minipage}
\end{table*}

The measured/model flux density ratios for both the nominal bias mode
and bright source mode data generally lie $\sim0.5$\% or $\sim1\sigma$
below unity, although a slightly larger systematic offset is seen in
the nominal mode data.  Given that the peak of the beam cannot be fit
to an accuracy better than $\sim1$\%, this performance is actually
very good.  We measured some systematic offsets of $\sim1$\% between
the median 350~$\mu$m ratios for the small scan maps and the ratios
for the large scan maps, but these offsets are $\sim1\sigma$.  They
may be the consequence of the minor differences in coverage of the
beam in the small and large scan maps.  Some large scan map ratios are
offset from unity by $>0.01$, but for any observation, only one
waveband deviates by this amount.

The small scan map observations were performed more frequently than
the large scan map observations and also do not include any notable
outliers, so we used the small scan map measured/model ratios to
examine long-term trends in the reproducability of the Neptune flux
density.  We examined this by fitting lines to the data and then
calculating the difference in the best fitting line between OD 100 and
OD 1450 (the approximate time range during which SPIRE was
operational).  Except for the nominal mode 250~$\mu$m data, all small
scan map ratios decrease by $\sim 0.7$\% between these ODs.  The
significance of this decrease is typically below the $3\sigma$ level.
Based on these data alone, it is unclear whether this is an issue with
the Neptune models, a long-term change in the sensitivity of SPIRE, or
some other misdiagnosed systematic effect.  Note that the trends in the
bright source mode data are dependent upon just three data points
before OD 400, although the similarity between the trends seen in the
nominal and bright source mode data imply that the ratios are changing
in data taken using both bias modes.

\subsection{Tests with Uranus data}

\begin{figure}
\epsfig{file=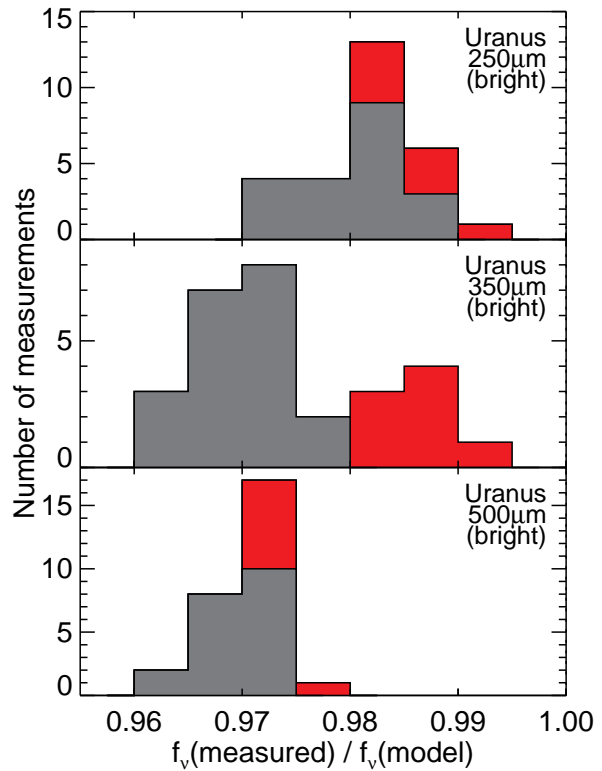}
\caption{Histograms of the measured to model flux densities for
  Uranus.  The grey data represent measurements made in small scan map
  mode, and the red data represent measurements made in large scan map
  mode.}
\label{f_hist_urn}
\end{figure}

\begin{figure}
\epsfig{file=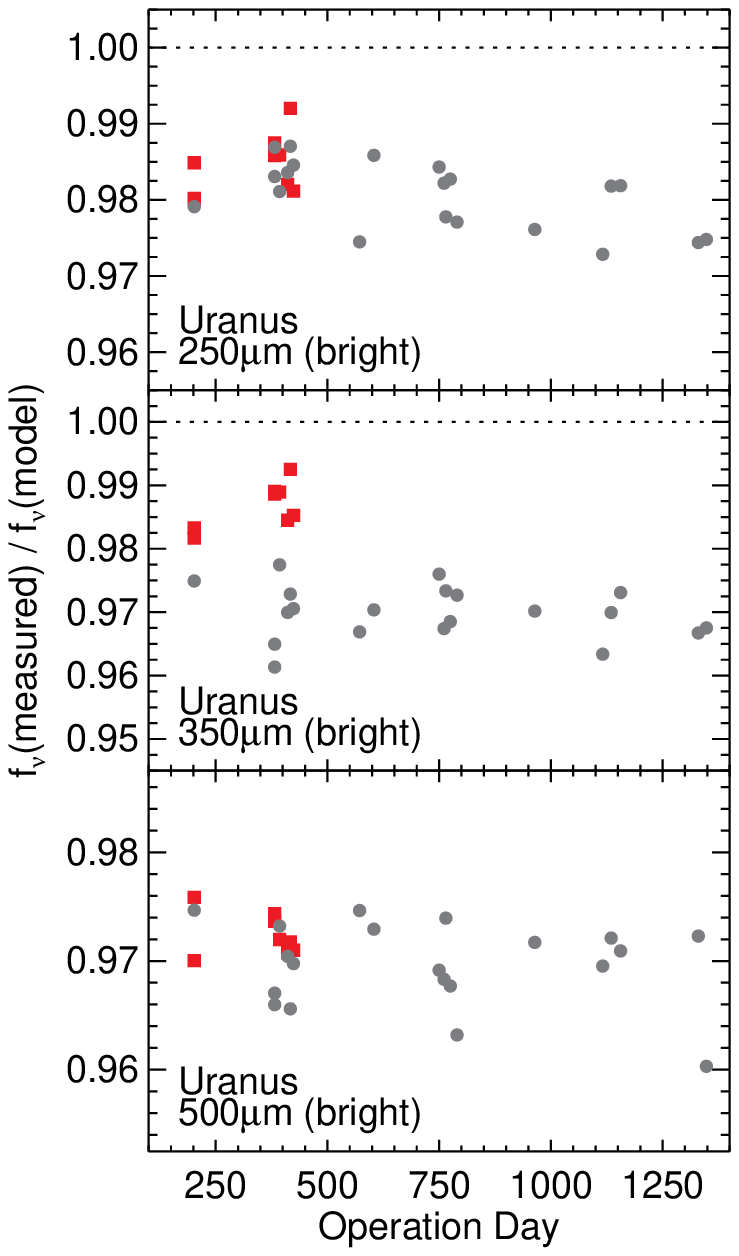}
\caption{Ratios of the measured to model flux densities for Uranus
plotted as a function of OD.  The grey circles are measurements made
in small scan map data, and the red squares are measurements made in
large scan map data.  The uncertainties from the fits are equivalent
to or smaller than the symbols in these plots.}
\label{f_test_urn}
\end{figure}

As with the Neptune data, we present statistics on the ratio of the
measured flux densities to the model flux densities for Uranus because
the distance to Uranus and hence the observed flux density will vary
over time.  The ratios are shown in
Figures~\ref{f_hist_urn}-\ref{f_test_urn} and Table~\ref{t_test_urn}.
The model flux densities are derived from the ESA-4 version of the
Uranus planetary atmosphere model (Orton, private communication).  We
did not use data from observation 1342233337 because of quality
control issues related to calculating the correct voltage offsets for
that specific observation.

\begin{table*}
\centering
\begin{minipage}{111mm}
\caption{Measured/model flux density ratios for Uranus}
\label{t_test_urn}
\begin{tabular}{@{}lcccc@{}}
\hline
Waveband &   \multicolumn{3}{c}{Measured/Model Flux Density Ratios} &
             Change in Ratios between\\
($\mu$m)&    All &             Small &                Large &
             OD 100 and OD 1450$^a$\\
\hline
250 &        $0.982\pm0.006$ & $0.982\pm0.006$ &      $0.985\pm0.004$ &
             $-0.010\pm0.003$\\    
350 &        $0.973\pm0.010$ & $0.970\pm0.006$ &      $0.987\pm0.004$ &
             $-0.003\pm0.004$\\
500 &        $0.971\pm0.004$ & $0.970\pm0.005$ &      $0.972\pm0.002$ &
             $-0.003\pm0.003$ \\
\hline
\end{tabular}
$^a$ These values are calculated by fitting lines to the
    measured/model ratios for the small scan map data (excluding the
    data point from OD 967) and then calculating the difference in the
    best fitting line between OD 100 and OD 1450.
\end{minipage}
\end{table*}

We systematically measure flux densities that are 2-3\% lower than the
model flux densities for Uranus, indicating that the Neptune and
Uranus models are consistent to within 3\%.  This lies within the 4\%
uncertainties of the Uranus and Neptune models.  A
statistically-significant difference is seen between the 350~$\mu$m
large and small scan map ratios, which again reflects differences in
the coverage of the beam peak.  Also, the observations on OD 967
(1342237551) produced a measurement/model ratio that is $\sim2$\%
lower in all three bands than the ratios measured in any other
observations.  The reason for this discrepancy is unclear, as the
observations are no different from any others in terms of the
observation set-up or quality.  If we exclude the OD 967 data from our
dataset, the standard deviation in the measured/model ratios for the
small scan map data decreases to 0.004, but the results are not
otherwise significantly affected.

Ignoring the OD 967 data, we see a 1.0\% decrease in the
measured/model ratio for the 250~$\mu$m data between OD 100 and OD
1450 (although this is only measured at the $\sim3\sigma$ level) but no
significant change in the 350 and 500~$\mu$m ratios.  The changes in
the ratios for Uranus and Neptune bright source mode observations are
similar for all three bands, which implies that it could be an
instrument-related effect, albeit an effect that is barely
statistically significant.

\subsection{Tests with Gamma Dra data}

\begin{figure}
\epsfig{file=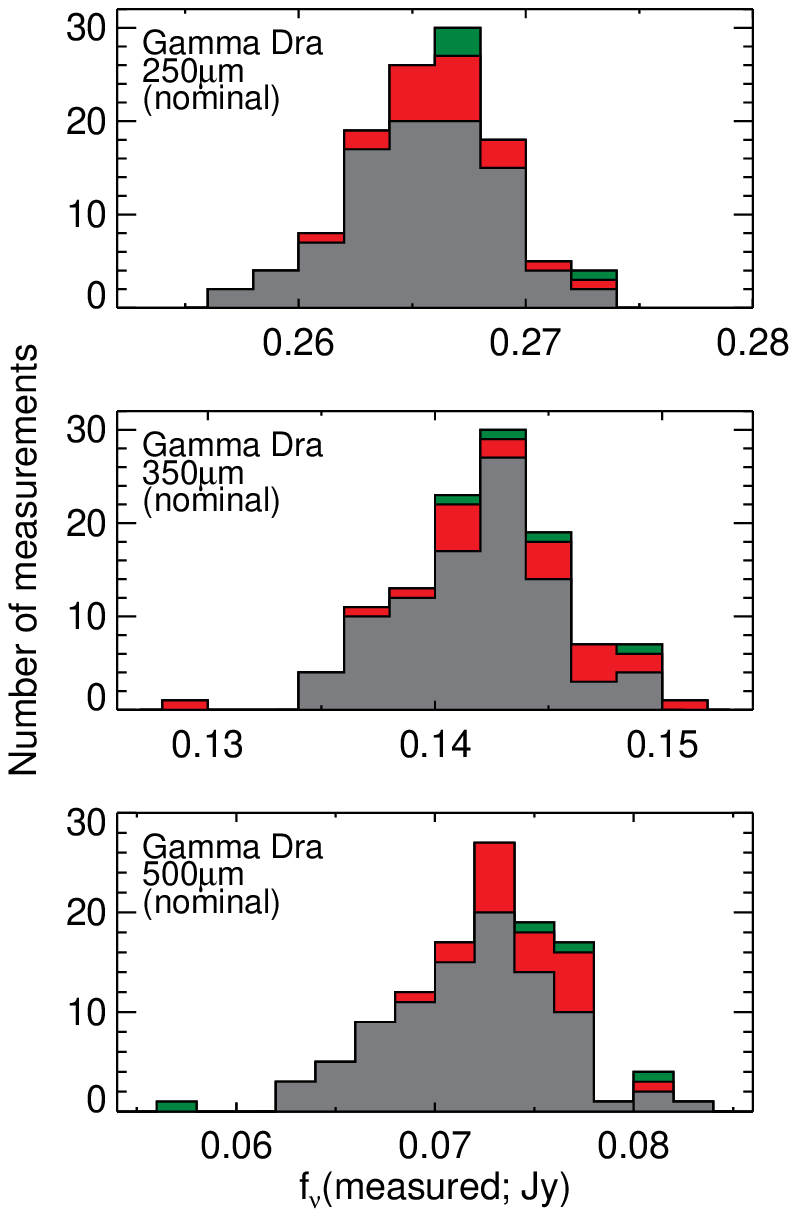}
\caption{Histograms of the measured to model flux densities for
  Uranus.  These flux densities do not include colour corrections.
  The grey data represent measurements made in small scan map data,
  the red data represent measurements made in large scan map data, and
  the green data represent measurements made in parallel mode map
  data.}
\label{f_hist_gammadra}
\end{figure}

\begin{figure}
\epsfig{file=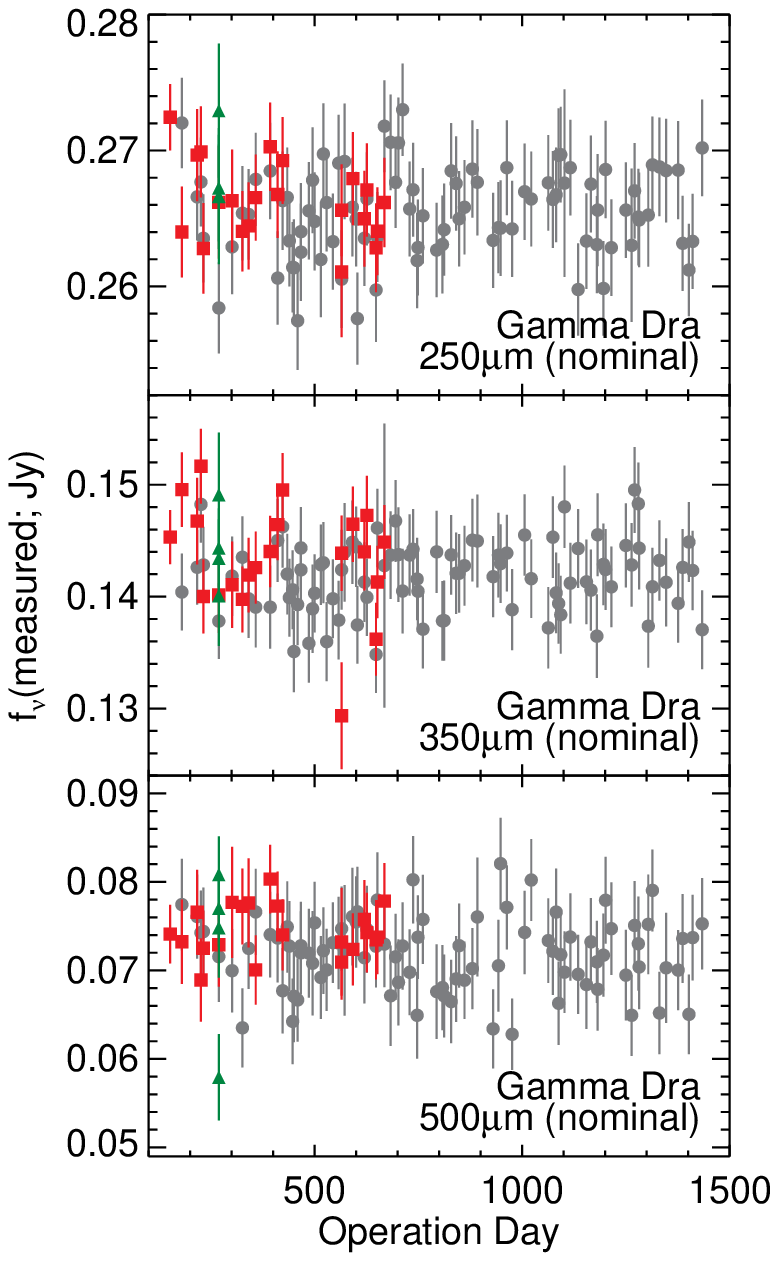}
\caption{Flux densities for Gamma Dra plotted as a function of OD.
These flux densities do not include colour corrections.  The grey
circles are measurements made in small scan map data, the red
squares are measurements made in large scan map data, and the green triangles
are measurements made in parallel mode map data.}
\label{f_test_gammadra}
\end{figure}

The flux density of Gamma Dra is not expected to vary over time, so we
report statistics on the flux densities themselves (with no colour
corrections) in Table~\ref{t_test_gammadra}, show histograms of the
flux densities in Figure~\ref{f_hist_gammadra}, and show the flux
densities as a function of time in Figure~\ref{f_test_gammadra}.  We
changed the timeline-based fitting method so that it would fit
circular two-dimensional Gaussian functions instead of elliptical
Gaussian functions, as tests with simulated sources with the same
brightness as Gamma Dra demonstrated that measurements based on
circular Gaussian functions produced slightly more accurate and more
precise results.  One of the observations (1342238335 from OD 989)
produced results with spuriously high uncertainties and is excluded
from this analysis.  The PSF was sparsely sampled in the parallel mode
observations using the fast scan rate, so we made several changes to
the technique used to fit Gaussian functions to the PSFs to optimise
it for these data\footnote[17]{For the parallel mode data taken at the
  fast scan speed, we processed the data with the wavelet deglitcher
  disabled; in version 10.0.620 of HIPE, this module was
  misidentifying Gamma Dra as a glitch.  Disabling the deglitcher
  resulted in excess noise in the background annulus, so in fitting a
  PSF to the data, we only measured a median signal in the background
  annulus data and fixed the background level during the fit rather
  than using the data in the background annulus in the fit and
  treating the background level as a free parameter.  We also fixed
  the FWHM of the PSF to the geometric mean values given by
  \citet{spire11}, which mitigated issues with the sparse sampling of
  the beam in these data.}

\begin{table*}
\centering
\begin{minipage}{128mm}
\caption{Measured flux densities for Gamma Dra}
\label{t_test_gammadra}
\begin{tabular}{@{}lccccc@{}}
\hline
Waveband &   \multicolumn{4}{c}{Flux Densities (Jy)$^a$} &
             Fractional Change in\\
($\mu$m)&    All &             Small &                Large &
             Parallel &
             Flux Density between\\
&            Scan Maps &       Scan Maps &            Scan Maps &
             Mode Maps &
             OD 100 and OD 1450$^b$\\
\hline
250 &        $0.266\pm0.003$ & $0.266\pm0.003$ &      $0.266\pm0.003$ &
             $0.267\pm0.003$ &
             $0.005\pm0.005$\\    
350 &        $0.142\pm0.004$ & $0.142\pm0.003$ &      $0.144\pm0.005$ &
             $0.144\pm0.004$ &
             $0.009\pm0.008$\\
500 &        $0.073\pm0.004$ & $0.072\pm0.004$ &      $0.074\pm0.003$ &
             $0.76\pm0.10$ &
             $-0.011\pm0.021$ \\
\hline
\end{tabular}
$^a$ These flux densities include no colour corrections.\\
$^b$ These values are calculated by fitting lines to the
    measured/model ratios for the small scan map data and then
    calculating the difference in the best fitting line between OD 100
    and OD 1450.
\end{minipage}
\end{table*}

Gamma Dra is much fainter than Neptune and Uranus, so the measurements
show more dispersion, particularly at 350 and 500~$\mu$m.  Unlike the
Neptune and Uranus data, no statistically-significant differences are
seen between the large and small scan map measurements.  However, the
0.5\% effects seen for Neptune would be undetectable in the Gamma Dra
data with its lower signal-to-noise ratio.  The Gamma Dra timeline
data exhibit relatively more scatter near the peak of the source than
the Neptune or Uranus timeline data, so it is also possible that minor
coverage differences between the small and large scan map measurements
do not significantly affect the function fit to the data.  The
parallel mode measurements show no statistically-significant
difference from the large and small scan map measurements, although
one of the fast scan speed parallel map measurements at 500~$\mu$m is
$\sim4\sigma$ lower than the median measured in other data (probably
because of the lower signal-to-noise ratio in these data and the
faintness of the source at 500~$\mu$m).

We did not measure any statistically significant change in any of the
data between OD 100 and OD 1450.  The Gamma Dra data may lack the
sensitivity needed to detect the $\sim0.7$\% decrease in detector
response that may be implied by the Neptune data.  However, the
measured increase of 0.5\% in the 250~$\mu$m data is $\sim2\sigma$
greater than the expected 0.7\% decrease.  This suggests that the
change in the Neptune and Uranus measured/model flux density ratios at
250~$\mu$m may actually be related to issues with modelling temporal
changes in flux densities for the planets, although the evidence for
this is tenuous.

\section{Summary of the assessment of the flux calibration}
\label{s_summary}

We have outlined the methods by which Neptune is used as the primary
flux standard for the {\it Herschel}-SPIRE photometer, including a detailed
assessment of the overall error budget associated with transferring
the Neptune calibration to an unknown point source.

The flux calibration for all individual bolometers has been thoroughly
assessed.  The relative uncertainties are typically $\sim0.5$\% for most
bolometers in both calibration modes.  However, because of the
problems with the truncated signal during the Neptune observations,
the uncertainties for some individual bolometers in the nominal bias
mode are $\sim1-5$\%.

The primary assessment of the flux calibration uncertainties for each
array as a whole is based on the Neptune data.  We were able to
measure the flux density of Neptune to within 1.5\% of the model flux
density in all three bands and using both voltage bias modes.  This
uncertainty includes both the systematic offset between the measured
and model flux densities and the $1\sigma$ dispersion in the
measurements.  As all Neptune data were used to calculate this
uncertainty, it encompasses any possible temporal changes in the
detector sensitivity during the mission and any variability in the
brightness of Neptune not accounted for by the models, although the
evidence for either is inconclusive.  The uncertainty also includes
the variations in measurements between different observing modes,
which is mainly a consequence of minor differences in the coverage. 
We therefore conclude that 1.5\% can be adopted as the relative
calibration uncertainty for the SPIRE photometer arrays.  The overall
error budget must also include the 4\% absolute uncertainty ascribed to
the Neptune model, and any statistical or other uncertainties
associated with a particular measurement.

\section*{Acknowledgments}

We thank the reviewer for the helpful comments on this paper. SPIRE
has been developed by a consortium of institutes led by Cardiff
Univ. (UK) and including: Univ. Lethbridge (Canada); NAOC (China);
CEA, LAM (France); IFSI, Univ. Padua (Italy); IAC (Spain); Stockholm
Observatory (Sweden); Imperial College London, RAL, UCL-MSSL, UKATC,
Univ. Sussex (UK); and Caltech, JPL, NHSC, Univ. Colorado (USA). This
development has been supported by national funding agencies: CSA
(Canada); NAOC (China); CEA, CNES, CNRS (France); ASI (Italy); MCINN
(Spain); SNSB (Sweden); STFC, UKSA (UK); and NASA (USA).  HIPE is a
joint development by the Herschel Science Ground Segment Consortium,
consisting of ESA, the NASA Herschel Science Center, and the HIFI,
PACS and SPIRE consortia.

{}

\label{lastpage}

\end{document}